# Two-Dimensional Halide Perovskites: Synthesis, Optoelectronic Properties, Stability, and Applications

Sushant Ghimire[a] and Christian Klinke*[a,b,c]

* Corresponding author

[a] Institute of Physics, University of Rostock, 18059 Rostock, Germany

[b] Department of Chemistry, Swansea University, Swansea SA2 8PP, U.K.

[c] Department "Life, Light & Matter", University of Rostock, Albert-Einstein-Strasse 25, 18059 Rostock, Germany

**Corresponding author:** christian.klinke@uni-rostock.de

Halide perovskites are promising materials for light-emitting and light-harvesting applications. In this context, two-dimensional perovskites such as nanoplatelets or Ruddlesden-Popper and Dion-Jacobson layered structures are important because of their structural flexibility, electronic confinement, and better stability. This review article brings forth an extensive overview of the recent developments of two-dimensional halide perovskites both in the colloidal and non-colloidal forms. We outline the strategy to synthesize and control the shape and discuss different crystalline phases and optoelectronic properties. We review the applications of two-dimensional perovskites in solar cells, light-emitting diodes, lasers, photodetectors, and photocatalysis. Besides, we also emphasize the moisture, thermal, and photostability of these materials in comparison to their three-dimensional analogs.

**TOC graphic**

2D halide perovskites show stable layered structures with interesting properties such as electronic confinement, energy funneling, exciton dissociation at crystal edges, and broad emission due to self-trapped excitons (STEs). These properties make 2D halide perovskites promising for solar cells, LEDs, lasers, photodetectors, and photocatalysis.

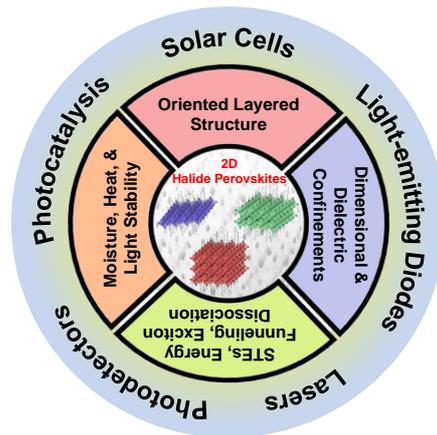

# Biography of Authors

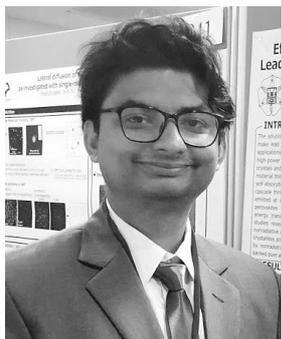

Sushant Ghimire obtained his M.Sc. degree in chemistry from Tribhuvan University (Nepal) in 2015. In 2016, he was awarded a Japanese Government (Monbukagakusho: MEXT) Scholarship for doctoral research. He received his Ph.D. degree in material science from Hokkaido University (Japan) in 2020. He is currently an Alexander von Humboldt Research Fellow at the University of Rostock (Germany). His current research interests include synthesis and optoelectronic properties of two-dimensional metal halide perovskites.

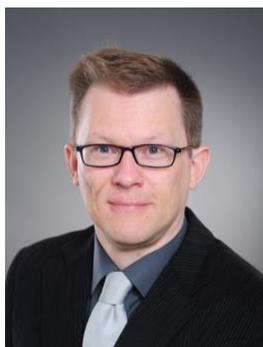

Christian Klinke studied physics at the University of Würzburg and the University of Karlsruhe (Germany) where he also obtained his diploma degree. In March 2000, he joined the Institute of Experimental Physics of the EPFL (Lausanne, Switzerland). Then from 2003 on, he worked as Post-Doc at the IBM T. J. Watson Research Center (Yorktown Heights, USA). In 2006, he became member of the University of Hamburg (Germany) and in 2007, he started as assistant professor at the University of Hamburg. In 2009, he received the German Nanotech Prize (Nanowissenschaftspreis, AGeNT-D/BMBF). His research was supported by an ERC Starting Grant and a Heisenberg fellowship of the German Funding Agency DFG. Since 2017 he is an associate professor at the Chemistry Department of the Swansea University and since 2019 full professor at the Institute of Physics of the University of Rostock. He is also member of the Interdisciplinary Department "Life, Light & Matter". His research concerns the colloidal synthesis of nanomaterials and the optoelectronic characterization of these materials.

1. Introduction

Halide perovskites have shown significant efficiencies in solar cells[1,2] and LEDs.[3,4] Photodetection,[5,6] X-ray scintillation,[7,8] lasing,[9,10] and photocatalysis[11,12] are other areas where these materials are flourishing. Halide perovskites are referred to as defect tolerant[13–15] which, along with the low exciton binding energy in their bulk, show long-range carrier diffusion and high carrier mobility.[16–18] A record-breaking diffusion length of 175 µm is already demonstrated in bulk methylammonium lead iodide (MAPbI$_3$) single crystal.[17] An unusually long photoluminescence (PL) lifetime has also been reported in the close-packed assembly of perovskite nanocrystals.[19,20] Further, these materials show high photoluminescence (PL) quantum yield (QY) and halide composition-dependent bandgap energy which is tunable in the entire visible spectrum.[9,21–23] Such unique optoelectronic properties make halide perovskites highly promising for light-harvesting and light-emitting applications.

Three-dimensional (3D) perovskites have ABX$_3$ structures with corner-sharing inorganic [BX$_6$]$^{4-}$ octahedra, where the B-site cation is Pb$^{2+}$, Sn$^{2+}$, or Ge$^{2+}$ and X$^-$ ions are halides (Cl$^-$, Br$^-$, or I$^-$). The A-site cation occupies the space formed by the corner-sharing inorganic octahedra to form the crystal lattice. The formation of a thermodynamically stable 3D perovskite lattice is guided by the Goldschmidt tolerance factor ($t = \frac{r_A + r_X}{\sqrt{2}(r_B + r_X)}$) and the octahedral factor ($\mu = \frac{r_B}{r_X}$), which correlate the ionic radii of A-site ($r_A$), B-site ($r_B$), and X$^-$ ($r_X$) ions. A stable ABX$_3$ perovskite structure is formed with Cs$^+$, MA$^+$ (methylammonium), or FA$^+$ (formamidinium) as A-site cations which result in the Goldschmidt tolerance factor in a narrow range of 0.81 to 1.[18,24] This limits the structural flexibility in 3D perovskites. The tunable optical properties in ABX$_3$ perovskites are governed by the band-edge electronic structure formed by the mixing of *s*- and *p*-orbitals of B-site cations and halides.[25] Despite having exceptional optoelectronic properties, 3D halide perovskites are less stable in a moist environment and under intense light irradiation which challenges their commercial use in photovoltaics and lighting devices.[26–32]

In this review, we discuss two-dimensional (2D) halide perovskites which include Ruddlesden-Popper (R-P)[33] or Dion-Jacobson (D-J)[34] layered structures and their colloidal quasi-2D nanoplatelets/nanosheets.[35,36] While the 2D R-P and D-J layered perovskites are (100)-oriented structures that are obtained by cutting the 3D $ABX_3$ perovskite along the (100) plane, various (111)- and (110)-oriented 2D halide perovskites are also known.[37,38] Besides, an interesting and new class of layered 2D halide perovskites called alternating cations in the interlayer space (ACI) perovskites have been reported in which the layered structure is stabilized by the alternate ordering of the guanidinium and methylammonium cations in the interlayer space.[39,40] These ACI perovskites show high crystal symmetry and are promising candidates for high-efficiency solar cells and LEDs.[41–43] In 2D R-P or D-J layered perovskites, the Goldschmidt tolerance factor is relaxed, and therefore they can accommodate a wide variety of bulky organic cations to obtain a thermodynamically stable structure with the desired optoelectronic properties.[44–46] In other words, 2D R-P or D-J layered perovskites are structurally more flexible than the $ABX_3$ types. Further, layered 2D halide perovskites have better properties than their 3D analogs in terms of stability and optoelectronics which are promising for energy-harvesting and light-emitting applications.[47,48] Therefore, a comprehensive review that provides insight into the synthesis, properties, stability, and applications of 2D halide perovskites is currently in demand. Here, we review the influence of temperature, organic ligands, spacer cations, and solvents on the dimensionality and crystal phases of 2D halide perovskites. Further, we discuss the effect of dimensional and dielectric confinements on optoelectronic properties. The applications of these materials in solar cells, LEDs, lasers, photodetector, and photocatalysis are summarized. We present an overview of the moisture, thermal, and photostability of the material. This review article covers a broad area of 2D halide perovskites. We believe it provides a good reference for students and researchers in understanding the fundamental properties of these classes of materials and advance further.

## 2. Crystal Structure

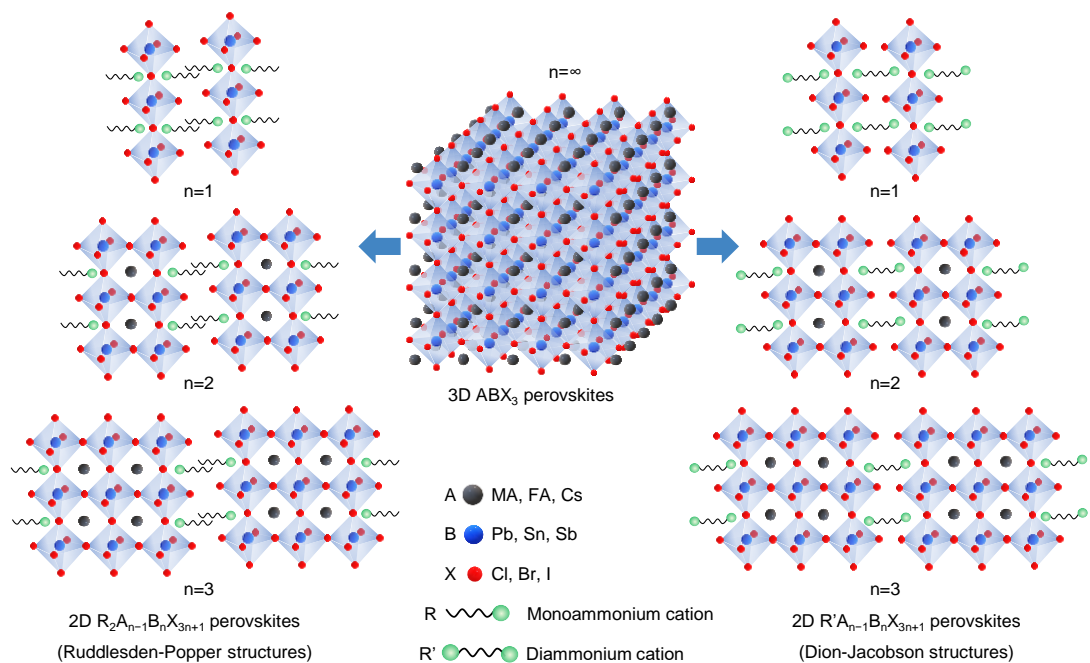

**Figure 1.** Crystal structures of 2D and 3D halide perovskites.

In 2D R-P or D-J layered perovskites, the $A_{n-1}B_nX_{3n+1}$ layer is sandwiched between the bulky organic cations, where n refers to the number of unit cells or thickness of the perovskite slab.[33,34,37,45,49–55] When n=1, a monolayered 2D structure is formed. Transformations to multilayered quasi-2D structures take place when n ≥2. The 2D R-P layered perovskite forms the $R_2A_{n-1}B_nX_{3n+1}$ structure, where monovalent organic cations ($R^+$) are hydrogen-bonded to the perovskite layers only at one side, whereas the inter-layer cations are interdigitated and interact with each other by van der Waals forces.[33,37,45,49–51,54] On the other hand, in 2D D-J layered perovskites, a single divalent organic cation ($R^{2+}$) is hydrogen-bonded at its two ends to the two different perovskite layers forming the $RA_{n-1}B_nX_{3n+1}$ structure.[34,45,52] Monoammonium organic cations such as *n*-butylammonium (BA),[33,54,56,57] *iso-* butylammonium (*iso*-BA),[56,57] *n*-pentylammonium (PA),[50,54] *n*-hexylammonium (HA),[50,54] *n*-octylammonium (OA),[58] phenylethylammonium (PEA),[46,59–61] and phenylmethylammonium (PMA) or benzylammonium[62,63] are used as organic spacers in 2D R-P layered perovskites. Similarly, (aminomethyl)piperidinium (AMP),[34,64–67] (aminomethyl)pyridinium (AMPY),[52] 1,4-phenylenedimethanammonium (PDMA),[68–70] and short-chain alkyldiammonium cations[71–73] are some examples of organic spacers in 2D D-J layered perovskites. Figure 1 distinguishes between the layered 2D and 3D halide perovskite structures. The flexibility, arrangement, and separation of inorganic stacking in layered 2D halide perovskites are controlled by the choice of organic spacer cations. The monoammonium spacer cations in 2D R-P layered perovskites allow the stacking layers to slide over each other, resulting in a staggered arrangement due to half a unit cell interlayer shift. On the other hand, the divalent organic cations increase the rigidity of the 2D D-J layered structure giving rise to an eclipsed arrangement of the stacking layers with smaller interlayer spacing.[45,52,54] Nevertheless, a 2D R-P layered structure is formed if such divalent organic cations are sufficiently long to tilt and allow the inorganic layers to slide in a staggered configuration.[53]

*2.1 Characterization*

In (100)-oriented 2D halide perovskites, the A-site cations are fully or partly replaced by the monodentate or bidentate organic ammonium cations to form R-P or D-J layered structures with different thicknesses of the inorganic layer (n=1, 2, 3, and so on). Unlike $ABX_3$ perovskites, the layered 2D halide perovskites have the inorganic $BX_6$ octahedra in two different environments, those which are on the surface next to the organic spacer layer, and the next ones which are at the middle surrounded by other $BX_6$ octahedra. Therefore, the crystal structure of a layered 2D halide perovskite is influenced on the one hand in a similar way to that of $ABX_3$ perovskites, and on the other hand in a unique way due to the presence of organic spacer cations.[33,34,52,54] The stacked inorganic layers in layered 2D halide perovskites give rise to the texture effect which is reflected in the corresponding XRD with the appearance of periodic ($h$00) diffraction peaks. For example, as shown in Figure 2a, $(BA)_2(MA)_{n-1}Pb_nI_{3n+1}$ perovskite with n=1, 2, 3, or 4 bears two, three, four, or five of the constantly spaced ($h$00) reflections, respectively below the (111) diffraction peak.[54] The periodicity of these peaks corresponds to the average spacing between the 2D halide perovskite layers. The diffraction peaks shift to lower 2$\theta$ angle values with an increase in the number of layers or n. Weidman et al. discussed the variation of spacing between the inorganic layers in $R_2(FA)_{n-1}Pb_nBr_{3n+1}$ perovskite nanoplatelets with different thickness (n=1, 2) and different ligands (R=BA, OA, BA/OA mixture), where the periodicity of the (*h*00) reflections increased with increasing the number of layers n or the ligand chain length.[74] A similar trend is observed in the case of 2D D-J layered structures (Figure 2 c-e).[52] Further, the presence of periodic (*h*00) peaks can also be used to confirm the purity of the layered 2D halide perovskites. For example, Klein et al. observed, in addition to the periodic (*h*00) diffraction peaks that correspond to a pure n-phase, the presence of extra diffraction peaks in the XRD patterns of layered MAPbX (X=Cl, Br, I) perovskite nanosheets which were from the minor fractions with different n or from the bulk perovskites.[36]

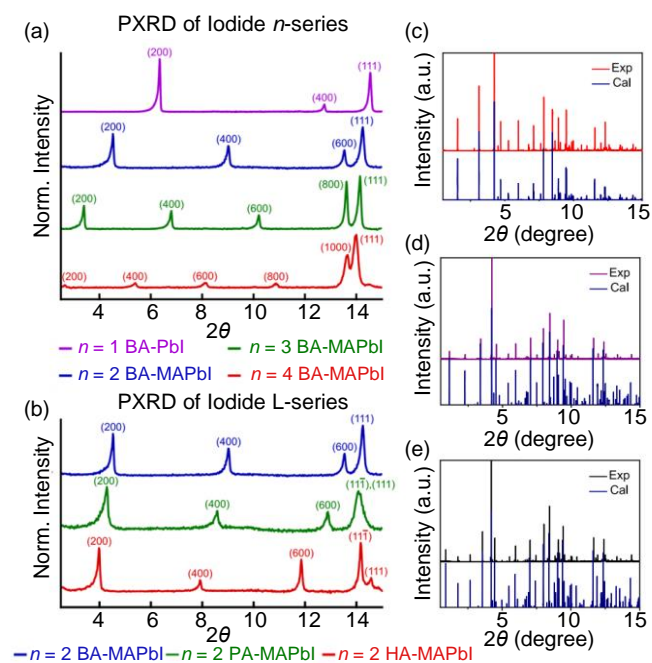

**Figure 2.** Characterization of 2D (a,b) R-P and (c-e) D-J layered perovskites using X-ray diffraction (XRD). Powder XRD patterns of (a) $(BA)_2(MA)_{n-1}Pb_nI_{3n+1}$ perovskite with n=1-4 and (b) $R_2MA_2Pb_2I_7$ perovskite with the different carbon chain lengths of organic spacers, where R=butylammonium (BA), pentylammonium (PA), and hexylammonium (HA). Reprinted with permission from ref. 54. Copyright (2019) American Chemical Society. (c-e) Powder XRD patterns of $(4AMPY)(MA)_{n-1}Pb_nI_{3n+1}$ perovskite with (c) n=2, (d) n=3 and (e) n=4. Reprinted with permission from ref. 52. Copyright (2019) American Chemical Society.

*2.2 Effect of Organic Spacer Cation*

The presence of different organic spacer cations greatly influences the crystal structure, symmetry, and thermodynamic stability of layered 2D halide perovskites.[45,50,54,75] As shown in Figure 2b, the increase in organic spacer cation chain length in 2D R-P layered perovskites from four (BA) to five (PA) and six (HA) carbons shift the periodic (*h*00) diffraction peaks to lower $2\theta$ angle which is attributed to an increase in the distance between the adjacent inorganic layers.[54] Also, $(BA)_2(MA)Pb_2I_7$ perovskite acquires an orthorhombic (C*mcm*) phase and $(PA)_2(MA)Pb_2I_7$ and $(HA)_2(MA)Pb_2I_7$ perovskites acquire a monoclinic (C2/*c*) phase at room temperature which is characterized by a splitting of the (111) diffraction peak in the latter two cases (Figure 2b).[54] Besides, the crystal symmetries in layered 2D halide perovskites are influenced by the number of inorganic layers n.[33,45,50,53,54,75] For example, $(BA)_2(MA)_{n-1}Pb_nI_{3n+1}$ perovskite series acquire the orthorhombic *Cc2m* or *Cmcm* space group for even and *C2cb* for odd number of inorganic layers.[45,54] Furthermore, the greater structural distortion/tilting in $PbX_6$ octahedra and lesser polarity in Pb-X bonds are observed at the surface closest to the organic spacer cation layer when compared to the middle.[50,54] Such octahedral distortions in 2D halide perovskites can occur as the deviation from the Pb−X bond length of $PbX_6$ octahedra.

Also, the out-of-plane tilting of the X-Pb-X bond angle and the in-plane tilting of the Pb-X-Pb bond angle from the ideal values of 90° and 180°, respectively contributes to the octahedral distortion. The Pb-X bond length deviation and the out-of-plane tilting are quantified in terms of bond length distortion ($\Delta d$) and bond angle variance ($\sigma$) of the of the PbX$_6$ octahedra, which are defined by the equations 1 and 2, respectively.[75]

$$\Delta d = \left(\frac{1}{6}\right) \sum_{i=6}^{6} \left[\frac{d_n^2 - d}{d}\right] \qquad 1,$$

where $d$ is the average Pb−X bond length and $d_n$ are the individual Pb−X bond lengths and

$$\sigma^2 = \sum_{i=1}^{12} \frac{(\theta_i - 90)^2}{11} \qquad 2,$$

where $\theta_i$ is the individual deviated X−Pb−X bond angle from the non-distorted 90°. In 2D halide perovskites, $\Delta d$ and $\sigma$ are larger for the R-P layered structures when compared to the D-J layered structures, indicating a larger octahedral distortion in the case of monoammonium organic cation-based 2D structures.[75] On the other hand, the organic spacer cation penetration depth into the plane of the axial halogens of the inorganic slab has a significant influence on the in-plane tilting of the PbX$_6$ octahedra.[45,73,76,77] This is accredited to the stress exerted on the inorganic layers by the different extent of electrostatic and hydrogen-bonding interactions among PbX$_6$ octahedra, A-site, and organic spacer cations. The octahedral distortion in the inorganic slab decreases by increasing the length of the organic spacer cations in layered 2D halide perovskites which is due to the minimization of the stress transmitted to the PbX$_6$ octahedra due to the presence of the long carbon chains.[54]

The position of the -NH$_3^+$ group and the presence of an aromatic ring in the organic spacer cation also influence the crystal structure of layered 2D halide perovskites through different distortions of the adjacent inorganic layers (Figure 3).[34,45,52,75] For example, in AMP-based 2D

D-J layered perovskites (Figure 3 a and c),[34,45] the presence of the $CH_3NH_3^+$ group at the 3rd position to nitrogen on the piperidine heterocyclic ring (3AMP) weakly hydrogen bonds the organic spacer cations to the terminal iodides on the surface. This has a small effect on the in-plane (equatorial) Pb-I-Pb angle distortion. On the other hand, 4AMP ($CH_3NH_3^+$ group at the 4th position to nitrogen on the piperidine heterocyclic ring) is hydrogen-bonded to the equatorial bridging iodide that is shared between the adjacent corner-sharing octahedra. This does not affect the out-of-plane (axial) tilting but increases the in-plane distortion. As the inorganic layer becomes thicker than n=2, the influence of organic spacer cation on the crystal structure decreases. A similar trend is observed in the case of AMPY-based 2D D-J layered perovskites (Figure 3 b and d).[45,52] However, the difference lies in the presence of the aromatic ring which decreases the interlayer spacing in AMPY-based layered 2D halide perovskites more than AMP-based analogs due to the aromaticity. Additionally, the 4AMPY-based 2D layered structure is more distorted compared to the 3AMPY-based structures. The strong hydrogen bonding of 4AMPY to the equatorial bridging iodide due to the ortho-para activation of the ring increases the in-plane distortion of stacking layers. On the other hand, the rigidity of the aromatic ring makes the 3AMPY cation slightly tilted when it is hydrogen-bonded to the stacking layers. Therefore, to accommodate a tilted 3AMPY cation, the stacking inorganic layers are slightly staggered which is in contrast to the 4AMPY-based 2D D-J layered perovskites that are perfectly eclipsed (Figure 3b).[52] Although the 2D D-J layered perovskites show smaller octahedral distortion when compared to the R-P layered structures, the former ones are thermodynamically less stable than the later ones, as shown in Figure 3e.[75] Also, the thermodynamic stability of layered 2D halide perovskites decreases as the thickness of the inorganic layer increases. It has been shown that the enthalpy of formation in the case of $(BA)_2(MA)_{n-1}Pb_nI_{3n+1}$ perovskite series becomes positive when n>5.[51]

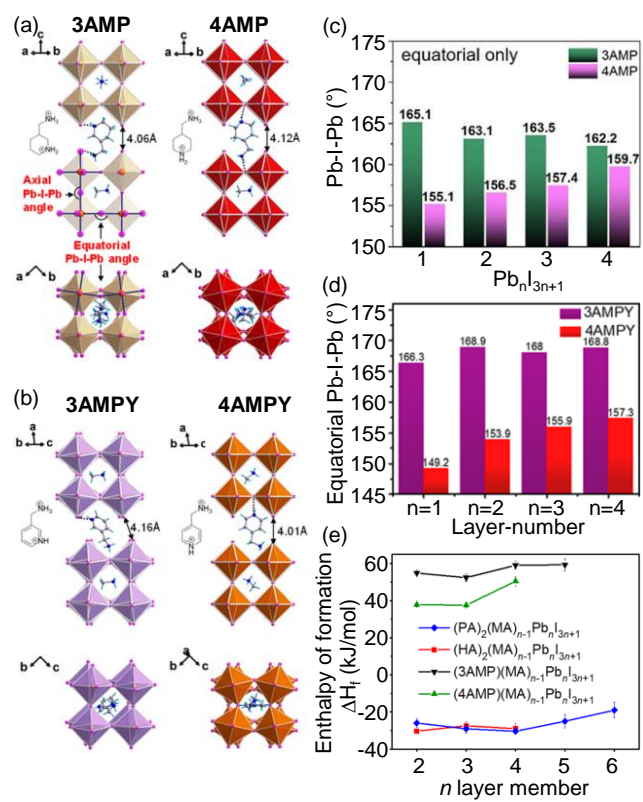

**Figure 3.** Octahedral distortion and thermodynamic stability of layered 2D halide perovskites. (a,b) Crystal structures of (a) AMP- and (b) AMPY-based 2D D-J layered perovskites. Reprinted with permission from ref. 45. Copyright (2021) American Chemical Society. (c,d) Equatorial Pb−I−Pb angles for (c) $(xAMP)(MA)_{n-1}Pb_nI_{3n+1}$ and (d) $(xAMPY)(MA)_{n-1}Pb_nI_{3n+1}$ (x=3, 4) perovskites at different inorganic layer thickness (n). Reprinted with permission from (c) ref. 34, copyright (2018) American Chemical Society and from (d) ref. 52, copyright (2019) American Chemical Society. (e) Enthalpy of formation of 2D R-P and D-J layered perovskites at different inorganic layer thickness (n). Reprinted with permission from ref. 75. Copyright (2021) American Chemical Society.

*2.3 Effect of Temperature*

Crystal phase transitions in halide perovskites are ubiquitous at different temperatures.[51,54,78–84] For example, bulk $MAPbI_3$ assumes a tetragonal (β) phase at room temperature which changes to a cubic (α) phase at the temperatures >310 K and an orthorhombic (γ) phase at temperatures <160 K.[78] Differential scanning calorimetry (DSC) and powder or single-crystal XRD techniques are used to quantify the phase transition temperature in perovskites (Figure 4 a and b).[81] In layered 2D halide perovskites, the phase transition temperature depends on the number of inorganic layers n and the carbon chain length/structure of the organic spacer cations.[50,51,54,82–84] Table 1 summarizes the temperature-dependent crystal phases and space groups of 4, 5, and 6 carbon chain alkylammonium-based layered 2D halide perovskites with different numbers of inorganic layers. All $(BA)_2(MA)_{n-1}Pb_nI_{3n+1}$ perovskites crystallize in the orthorhombic phase at room temperature, while $(PA)_2(MA)_{n-1}Pb_nI_{3n+1}$ and $(HA)_2(MA)_{n-1}Pb_nI_{3n+1}$ adopt the monoclinic phase at room temperature and the orthorhombic phase only at higher temperatures. An exception is $(HA)_2PbI_4$ perovskite which is orthorhombic at room temperature.[54] Interestingly, the phase transition temperature for an orthorhombic phase in BA-based 2D R-P layered perovskites is independent of n, whereas thickness-dependent phase transition temperatures are reported for the low-temperature triclinic phase of this material.[51,54] Also, in the case of PA- and HA-based 2D R-P layered perovskites, the phase transition temperature increases with an increase in the thickness of the inorganic layer. The transition from room temperature to the high or low-temperature phase is considered as a first-order (melting/freezing) transition of the alkylammonium chains separating the inorganic layers (Figure 4).[54,81] Such phase transitions in 2D halide perovskites above (or below) the room temperature is accompanied by an increase (or decrease) in the unit cell volume which is caused by the thermal expansion (or contraction) of the lattice. This shifts the adjacent inorganic layer relative to each other, changes the orientation (corrugation angle) of the organic spacer cations with respect to each other and the surface plane, and distorts the inorganic octahedra (Figure 4 c and d).[50,81] Mitzi and coworkers have shown that the branching of the alkyl chain near the ammonium group remarkably decreases the melting transition temperature in layered 2D halide perovskites.[85,86] This is attributed to the steric effect of the branched organic spacer cation on the electrostatic interaction with the inorganic layers. They also showed that the branched organic spacer with a long alkyl chain length additionally influences the van der Waals interaction between adjacent stacked layers.[86] Such temperature-dependent variation of crystal structures and lattice size in 2D halide perovskites can influence their optoelectronic properties and applications.[87–89]

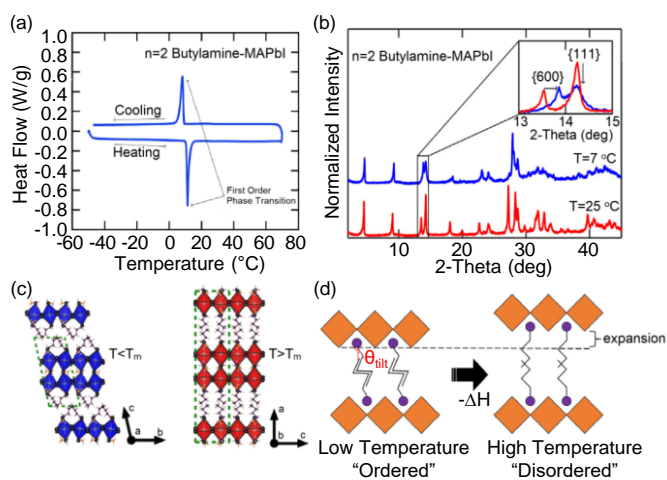

**Figure 4**. Crystal phase transition in layered 2D halide perovskites. (a) DSC heating−cooling curve revealing reversible first-order phase transition in $(BA)_2(MA)Pb_2I_7$ perovskite crystal. (b) Corresponding powder XRD above (red) and below (blue) the phase transition temperature, $T_m$, and inset showing the contraction of (600) intersheet spacing and broadening of (111) quasi-vertical plane. (c) Crystal structure of $(BA)_2(MA)Pb_2I_7$ perovskite below (left) and above (right) $T_m$. (d) Schematic representation of the observed phase transition. The corrugation tilt angle is shown schematically (red) as the angle between the organic chains and the normal vector of the inorganic sheet. Reprinted with permission from ref. 81. Copyright (2019) American Chemical Society.

**Table 1.** Crystal structure and symmetry of $R_2MA_{n-1}Pb_nI_{3n+1}$ (R=BA, PA, and HA; n=1 to 4) perovskites at different temperatures (T). RT refers to room temperature. This table is based on the references 33, 50, 54, and 82

| \multicolumn{9}{c}{$(BA)_2MA_{n-1}Pb_nI_{3n+1}$} |
|---|---|---|---|---|---|---|---|---|
| n | 1 | 2 | | 3 | | 4 | |
| T (K) | ~RT | 186-283 | >283~RT | 156-282 | >282~RT | 136-283 | >283~RT |
| Crystal system | Orthorhombic | Triclinic | Orthorhombic | Triclinic | Orthorhombic | Triclinic | Orthorhombic |
| Space group | Pbca | P$\bar{1}$ | Cmcm or Cc2m | P$\bar{1}$ | Cmca or C2cb | P$\bar{1}$ | Cmcm or Cc2m |
| \multicolumn{9}{c}{$(PA)_2MA_{n-1}Pb_nI_{3n+1}$} |
| n | 1 | | 2 | | 3 | | |
| T (K) | ~RT | >320 | ~RT | >351 | ~RT | >353 | |
| Crystal system | Monoclinic | Orthorhombic | Monoclinic | Orthorhombic | Monoclinic | Orthorhombic | |
| Space group | P2$_1$/a | Pbca | Cc or C2/c | Cmcm or Cc2m | Pc | C2cb | |
| \multicolumn{9}{c}{$(HA)_2MA_{n-1}Pb_nI_{3n+1}$} |
| n | 1 | | | 2 | | 3 | | 4 | |
| T (K) | <268 | ~RT | >355 | >371 | ~RT | >369 | ~RT | >376 | ~RT | >396 |
| Crystal system | Monoclinic | Orthorhombic | Orthorhombic | Tetragonal | Monoclinic | Orthorhombic | Monoclinic | Orthorhombic | Monoclinic | Orthorhombic |
| Space group | P2$_1$/a | Pbca | | | Cc | Cc2m | Pc | C2cb | Cc | Cc2m |

### 3. Crystal Growth

Crystallization of halide perovskites involves the mechanism of nucleation and growth.[90] In wet chemical synthesis, supersaturation of precursors in a solution starts the nucleation of perovskite crystals. The rate of nucleation depends on the degree of supersaturation, temperature, and surface free energy. The nucleation stops below a certain critical supersaturation concentration of the reacting precursors. The resulting perovskite nuclei then grow by the diffusion of precursors from the solution to the nuclei surface and reaction therein. The final shape/dimensionality of the halide perovskites is determined by the symmetry/structure of the nuclei. The symmetry of the nuclei can be preserved or systematically broken to obtain the material of the desired shape.[36,91–94] The interplay between different factors such as temperature, reaction time, precursors, ligands, solvents, and so forth influences the symmetry and growth of the perovskite nuclei where the size of the thus grown crystals is either bulk or confined to the nanoscale.[33,91,95–97] Figure 5 shows the microscopic images of 2D halide perovskites of different sizes and types. The influence of these parameters on the size and shape of halide perovskites are discussed below.

*3.1 Bulk crystal growth*

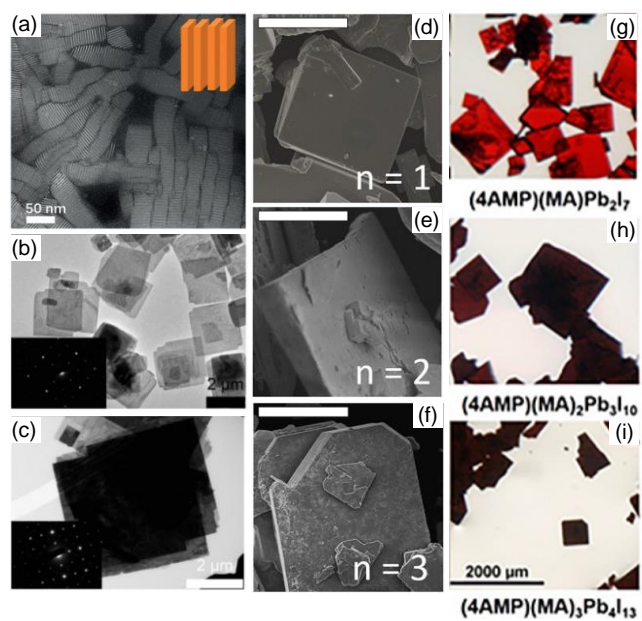

**Figure 5.** Microscopic characterization of 2D halide perovskites. (a) High-angle annular dark-field scanning transmission electron microscopic image of $CsPbBr_3$ perovskite nanoplatelets. Reprinted with permission from ref. 94. Copyright (2019) Wiley-VCH. (b,c) Transmission electron microscopic images of (b) MAPbI perovskite nanosheets with n≤4, and (c) MAPbI nanosheets with n>4. Insets show the corresponding selected area electron diffraction patterns. Reprinted with permission from ref. 36. Copyright (2019) American Chemical Society. (d-f) Scanning electron microscopic images of $BA_2MA_{n-1}Pb_nI_{3n+1}$ 2D R-P layered perovskite with (d) n=1, (e) n=2, and (f) n=3 (scale bar is 200 μm). Reprinted with permission from ref. 33. Copyright (2016) American Chemical Society. (g-i) Optical microscopic images of $(4AMP)MA_{n-1}Pb_nI_{3n+1}$ 2D D-J layered perovskite with (g) n=2, (h) n=3, and (i) n=4. Reprinted with permission from ref. 34. Copyright (2018) American Chemical Society.

Following the first report on the solution-phase crystallization of $(BA)_2(MA)_{n-1}Sn_{3n+1}$ perovskites by Mitzi et al.,[55] bulk crystals of homologous series of 2D halide perovskites with R-P or D-J layered structures are prepared by the slow cooling of their precursors in hydrohalic acid (Figure 5 d-i).[33,34,50–54,77,82–84] The thickness of perovskite layers is tuned systematically by adjusting the ratio between the bulky organic spacer cation and small A-site cation where the spacer cation acts as a limiting reagent. The use of spacer cations as limiting reagents is particularly important to obtain layered 2D halide perovskites of desired thickness.[33] In a typical synthesis of $(BA)_2(MA)_{n-1}Pb_nI_{3n+1}$ perovskites, the stoichiometric amounts of $PbI_2$ and MAI are dissolved in HI solvent and refluxed with half the stoichiometric amount of BAI. Upon cooling to ambient conditions, 2D R-P layered perovskites with the desired thickness or n crystallize (Figure 5 d-f). However, it is more challenging to synthesize the layered 2D halide perovskites of n>1 with diammonium spacer cations, where the crystallization of the thicker layers was prevented by the precipitation of monolayered 2D perovskites with n=1.[53] The low solubility of n=1 2D halide perovskites in HI is caused by the strong hydrogen bonding ability of the diammonium cations when compared to the monoammonium cations. The strong electrostatic interaction between the $-NH_3^+$ group of the organic spacer cations and $H_2O$ also results in the precipitation of undesirable hydrate (light-yellow) phases when the solution is very dilute or quickly cooled down.[45,52] Kanatzidis and coworkers successfully synthesized 2D D-J layered perovskites with n>1 using a nearly saturated solution of $Pb^{2+}$ and $MA^+$ in HI at its boiling point following a step-cooling.[52] At first, the precursor solution with organic spacer cations was cooled to a temperature just below the boiling point of HI (125 °C) and kept there for several hours under reflux to prevent the precipitation of an undesirable light-yellow phase. It was then slowly cooled to room temperature which yielded the 2D D-J layered perovskites of desired thickness.

Antisolvent vapor-assisted crystallization (AVC) can also be used for the synthesis of monolayer thick 2D halide perovskite bulk crystals.[45,88,98,99] This method is more common in the synthesis of 3D halide perovskites, where the micro to millimeter size crystals are formed by the slow vapor diffusion of a highly volatile antisolvent such as chlorobenzene or dichloromethane (DCM) into the precursors dissolved in dimethyl formamide (DMF), γ-butyrolactone, or dimethylsulfoxide (DMSO).[18] The antisolvents induce supersaturation by reducing the solubility of the perovskite phase in the solution which results in crystallization. In a typical synthesis, the stoichiometric amounts of the precursors $RNH_3X$ and $PbX_2$ (where $RNH_3$ is organic spacer cation and X is Br or I) are dissolved in the solvents such as DMF or γ-butyrolactone. This precursor solution is poured into a small container which is placed inside a larger container filled with DCM or chlorobenzene as the antisolvent. The whole system is covered and allowed to stand undisturbed for many hours to a few days which results in the crystallization of the layered 2D halide perovskites.[88,98,99] The AVC method can be further modified as the antisolvent vapor-assisted capping crystallization (AVCC) for the fast crystallization, typically within a few tens of minutes, of the layered 2D halide perovskites.[98] Here, the precursor solution is sandwiched between the two glass slides which are kept inside a larger container, and on the top of it, a small container filled with the antisolvent is placed. The AVCC method is beneficial over the AVC since it can result not only in the fast crystallization but also forms the multilayered 2D halide perovskites with n>1,[100] which is unlikely by the AVC. Nevertheless, the strong coordinating solvents can prevent the crystallization of the layered 2D halide perovskites which is due to the formation of the highly solvated precursor ions.[45]

*3.2 Growth on a substrate*

Halide perovskites can be grown as atomically flat 2D sheets directly from the evaporation of their precursor solution on a substrate such as silicon or glass.[59,101–106] The rate of solvent evaporation greatly influences the kinetics of crystal growth on the substrate. In turn, the polarity of solvents in which precursors are dissolved, the solvent volume ratio, the crystallization temperature, and the nature of the substrate surface affect the overall quality and thickness of the 2D halide perovskite structures. The deposition of the precursor solution on the substrate is carried out by drop-casting, spin-coating, chemical vapor deposition (CVD), or spray coating.[18] Our discussion here on the growth of 2D halide perovskites on a substrate is focused on drop-casting and spin-coating.

Yang and co-workers demonstrated the growth of single-crystalline $(BA)_2PbBr_4$ perovskite with single to few unit-cell thicknesses by drop-casting the precursor solution on a $Si/SiO_2$ substrate at a moderate temperature.[101] The thickness of the 2D perovskite sheets was controlled by using DMF-chlorobenzene-acetonitrile as a ternary co-solvent to dissolve the precursors. While chlorobenzene reduces the solubility of halide perovskites in DMF and promotes crystallization, fast evaporation of acetonitrile induces the formation and growth of ultrathin 2D perovskite sheets. Following this method, Chen et al. synthesized ultrathin $(BA)_2(MA)_{n-1}Pb_nBr_{3n+1}$ perovskite layers by doping the $(BA)_2PbBr_4$ sheets with MABr.[102] The thickness was tuned from n=2 to n=∞ by increasing the reaction time from 2 to 60 min. The mechanism of such structural transformation involves the intercalation of $MA^+$ cations between the $(BA)_2PbBr_4$ perovskite layers, followed by the release of $C_4H_9NH_3Br$ and simultaneous merging of $[PbBr_6]^{4-}$ octahedra. Spin-coating is another efficient technique to produce uniform and high-quality films of 2D halide perovskites.[59,105,107] This method is mostly used for depositing perovskite layers for device fabrication such as solar cells and LEDs. Here, the precursors dissolved in DMF or DMSO are spin-coated on glass/ITO substrates in a single step or multiple steps followed by ambient drying or annealing to obtain layered 2D halide perovskites. Karunadasa and coworkers showed the one-step spin coating of high-quality films of $(PEA)_2(MA)_2Pb_3I_{10}$ perovskite which acted as an excellent light-absorbing layer in the thus fabricated perovskite solar cells.[59] In addition to the stoichiometric ratio of precursors and size of organic cations, crystallization dynamics during the film growth largely determine the selective thickness of these 2D materials.[105,106] For example, the 2D halide perovskites obtained by the fast-crystallization of stoichiometric precursors in DMF through spin coating followed by the immediate high-temperature annealing do not always contain the stacking layers with purely one value of n.[105] They may contain a mixture of 2D layers with different n. On the other hand, a slow crystallization process rendered by the complexation of $Pb^{2+}$ with additives and low initial annealing temperatures can lead to the growth of 2D perovskite layers with purely one value of n.

*3.3 Hot-injection method*

The hot-injection method is a particular type of colloidal synthesis of monodispersed semiconductor nanostructures.[108] The burst nucleation of nanocrystals takes place when a precursor solution is injected into the hot solution of another precursor in the presence of ligands. The reaction is stopped after a certain time by sudden cooling of the reaction mixture to obtain quantum dots (QDs) or nanocrystals. Protesescu et al. demonstrated the hot-injection synthesis (here onwards standard hot-injection method) of $CsPbX_3$ (X=Cl, Br, I) perovskite nanocrystals at high temperatures (150 to 200 °C).[109] They employed cesium oleate and lead (II) halide as precursors, oleic acid and oleylamine as ligands, and octadecene as a high boiling solvent for the synthesis of 4-15 nm-sized $CsPbX_3$ perovskite nanocrystals. A very short reaction time of 30 s selectively resulted in the crystallization of cube-shaped nanocrystals under given conditions. Following these early reports, different modifications were made to the hot-injection method to obtain 2D or quasi-2D structures.[36,95–97,110–112] For example, Bekenstein et al. successfully showed the synthesis of quasi-2D $CsPbX_3$ perovskite nanoplatelets with the thickness varying from 1 to 5 unit cells.[110] Here, a lower reaction temperature (90-130 °C) than the standard hot-injection method resulted in the crystallization of nanoplatelets. The lower temperature (90 °C) slowed down the reaction kinetics and allowed the organic ligands to assemble into mesostructures inside which thin nanoplatelets were grown by the breaking of the crystals' inherent cubic symmetry. Shamsi et al. reported the formation of $CsPbBr_3$ perovskite nanosheets over a wide range of temperatures (50 to 150 °C) by using short-chain (octanoic acid and octylamine) and long-chain (oleic acid and oleylamine) ligands.[111] By increasing the ratio of short to long-chain ligands, they obtained larger and thinner nanosheets, while a longer reaction time (> 5 min) resulted in the aggregation of the products. The composition tuning of these nanosheets was possible through the post-synthesis halogen exchange. Klein et al. utilized pre-synthesized nanosheets of $PbX_2$ (X=Cl, Br, I) as templates in a hot-injection synthesis to obtain 2D MAPbX perovskite nanosheets (Figure 5 b and c).[36] The obtained nanosheets were ranging from 50 nm to 8 μm in their lateral size, and the thickness was varying from bulk (n≥4) to monolayer (n=1).

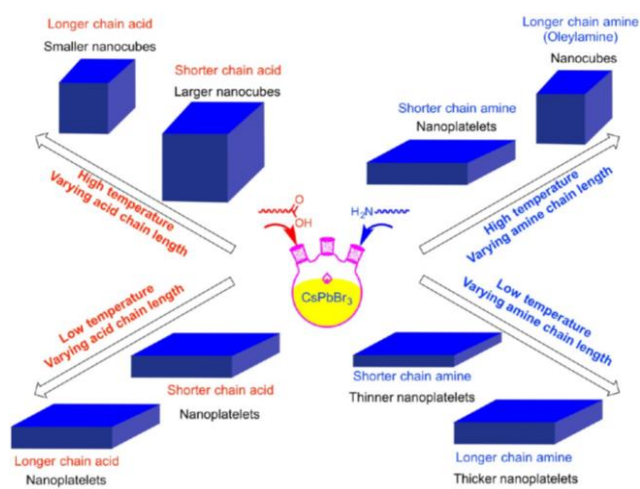

**Figure 6**. Effect of ligand chain length and temperature on the shape and size of halide perovskite nanocrystals synthesized using the hot-injection method. Reprinted with permission from ref. 96. Copyright (2016) American Chemical Society.

Hot-injection synthesis of halide perovskites can involve binary ligands, which are composed of aliphatic carboxylic acids and primary aliphatic amines. The precise regulation of chain length or concentration of these ligands can induce symmetry breaking, resulting in the growth of 2D or quasi-2D perovskite nanoplatelets. Pan et al. studied the effect of ligand chain length and temperature on the morphology of $CsPbBr_3$ perovskite nanocrystals (Figure 6).[96] While the hot-injection at higher temperature (170 °C) using shorter-chain carboxylic acids (dodecanoic, octanoic, or hexanoic acids) and a longer chain aliphatic amine (oleylamine) resulted in the crystallization of perovskite nanocubes, the use of a longer-chain carboxylic acid (oleic acid) and shorter chain amines (dodecyl-, octyl-, or hexylamine) at higher or lower temperatures resulted in the formation of nanoplatelets. Moreover, the thickness of the nanoplatelets was precisely controlled by decreasing the temperature to 140 °C and using the aliphatic amines with 6 or 8 carbon chain lengths. Manna and coworkers discussed the role of acid-base equilibria in directing the growth of perovskite nanoplatelets during the hot-injection synthesis.[95] The equimolar oleic acid and oleylamine ligands in a nonpolar solvent remain in equilibrium with the oleylammonium cation and oleate anion, which is exothermic. At higher concentrations of oleic acid or lower temperatures, the acid-base equilibrium shifts forward, increasing the concentration of oleylammonium ions. Such oleylammonium-rich conditions resulted in the growth of quasi-2D $[RNH_3]_2[CsPbBr_3]_{n-1}PbBr_4$ perovskite nanoplatelets. By using strong acids such as benzylsulfonic acid or hexanoic acid, they obtained perovskite nanoplatelets even at a lower concentration of acid. Moreover, by increasing the ratio of (oleylammonium ion)/$Cs^+$, which is by decreasing the concentration of $Cs^+$, they successfully synthesized the thickness-controlled perovskite nanoplatelets even at a higher temperature of 190 °C, which otherwise gives only the cube-shaped nanocrystals in the standard hot-injection synthesis.[95]

*3.4 Precipitation Method*

Colloidal 2D halide perovskites have also been conveniently prepared by precipitation, which is either by adapting the method reported by Schmidt et al.[113] or by using the ligand-assisted reprecipitation (LARP) technique first demonstrated by Zhang et al.[21] In the Schmidt method, precursor solutions of MAX and PbX$_2$ in DMF are added to a ligand mixture of oleic acid and the medium to long-chain alkyl ammonium halide in octadecene at 80 °C. Precipitation of nanocrystals is rendered by the subsequent addition of acetone. On the other hand, LARP involves the precipitation of perovskite nanocrystals at room temperature directly from the addition of precursors and ligands dissolved in polar solvents such as DMF, DMSO, or γ-butyrolactone into a non-polar solvent such as toluene. Following the Schmidt method, Tyagi et al. synthesized MAPbBr$_3$ perovskite nanocrystals at 80 °C by employing OABr as ligand.[114] 2D halide perovskites were isolated by purification of the reaction mixture through dilution, sonication, and filtration to remove large particles and selective precipitation of nanoplatelets using acetone. However, for the systematic control over the thickness of the nanoplatelets, stoichiometric variation of the ligand to precursor ratio is necessary. Sichert et al. systematically varied the ratio of OABr to MABr and obtained 2D halide perovskite nanoplatelets with thickness n=1 to n=∞.[35] Similarly, Akkerman et al. employed the Schmidt method to obtain CsPbBr$_3$ perovskite nanoplatelets with thicknesses ranging from 3 to 5 unit cells.[115] They used different polar solvents such as isopropyl alcohol, ethanol, and acetone to precipitate the nanocrystals. While isopropyl alcohol and ethanol resulted in large-sized particles due to the fast crystallization, the shape-controlled perovskite nanoplatelets were selectively precipitated by acetone.

The ligand-assisted reprecipitation (LARP) method is considered as one of the most versatile methods for the synthesis of 2D halide perovskites where the room temperature condition is ideal for the slow growth kinetics.[22,74,94,116–124] The crystallization process in LARP is caused by the supersaturation induced by the solubility change with solvent mixing. Tisdale and coworkers demonstrated the synthesis of colloidal lead and tin halide perovskite nanoplatelets, R$_2$A$_{n-1}$B$_n$X$_{3n+1}$ (A=Cs, MA, or FA, B=Pb or Sn, R=BA or OA, and X=Cl, Br, or I), with lateral sizes up to 1 $\mu$m and thicknesses down to one to two monolayers by the reprecipitation method.[74,123] The thickness of perovskite nanoplatelets was controlled by tuning the stoichiometric ratio of precursors RX, AX, and $BX_2$ in DMF. When the precursor stock solution was added to toluene at room temperature, layered 2D halide perovskite nanoplatelets of desired thickness were obtained. Nevertheless, the tin-based perovskite nanoplatelets were stable only under an inert atmosphere.[74] The thickness of the perovskite nanoplatelets synthesized via the LARP method can be controlled by the choice of the ligands or their concentrations, precursor ratio, and antisolvents. For example, Levchuk et al. varied the ratio of oleic acid and oleylamine ligands to obtain quantum-confined $MAPbX_3$ (X=Cl, Br, I) perovskite nanoplatelets with 2 to 5 monolayers.[117] While the use of chloroform as antisolvent resulted in thicker nanoplatelets, thinner ones were precipitated with toluene or a toluene/chloroform mixture. This is attributed to the different dielectric constants of the solvent and the antisolvent which allows chloroform instead of toluene to mix properly with DMF. As a result, nucleation kinetics in the case of chloroform become faster, and thicker perovskite nanoplatelets are formed. Nonetheless, either no precipitation of $FAPbX_3$ perovskite nanoplatelets or the formation of large particles and aggregates resulted when toluene was used as the antisolvent by the same group in their other work.[22] In this case, only pure chloroform worked well. Cho and Banerjee et al. carried out an extensive study on the dimensional control of $CsPbBr_3$ perovskite nanocrystals prepared from the ligand-mediated reprecipitation at room temperature.[118] A change in the shape of $CsPbBr_3$ perovskite nanocrystals from 3D to quasi-2D, 2D monolayer, and 0D structures was observed by increasing the molar ratio of the Pb precursor to different monodentate and bidentate amine ligands. Typically, ultrathin perovskite nanoplatelets were obtained at a Pb/n-octylamine ratio of 1:2, and the nanoplatelets with 3 to 4 monolayer thickness and large lateral dimensions were formed at the molar ratio of 1:4. High concentrations of coordinating amine ligands change the monomer supersaturation by binding with more $Pb^{2+}$ precursor ions. This resulted in the anisotropic growth of perovskite nanocrystals into 2D structures. On the other hand, at a very high concentration of amine ligands, these 2D halide perovskites were transformed into 0D $Cs_4PbBr_6$ perovskite nanocrystals which was caused by the leaching of $PbBr_2$ from the nanoplatelets. Recently, Polavarapu and co-workers demonstrated a modified LARP synthesis of formamidinium- or cesium-based halide perovskite nanoplatelets by tuning the molar ratio of monovalent ($FA^+$ or $Cs^+$) to divalent ($Pb^{2+}$) cations (Figure 5a).[94] While a decrease in the monovalent to divalent cation ratio resulted in the crystallization of 2D nanoplatelets, an increase in this ratio formed 3D nanocubes.

*3.5 Exfoliation*

Exfoliation is a top-down technique for synthesizing the 2D halide perovskites.[101,125–132] Scotch tape is used to obtain atomically thin 2D structures through repeated peeling of a bulk perovskite crystal by adhering it between the adhesive sides of two scotch tapes.[101,126,127] The thus obtained 2D halide perovskite layers are transferred onto a substrate by pressing the scotch tape against it. Although this method for obtaining 2D halide perovskites is straightforward, precise control over the thickness of inorganic layers is often difficult. 2D halide perovskites obtained from the scotch tape method are mechanically brittle,[101] and their stability and PL QY can be compromised due to the fractured surface caused by the mechanical force with large number of dangling bonds and surface defects. Recently, ligand-assisted ball milling is demonstrated by Yun et al. for the mechanical exfoliation of perovskite crystals to obtain quasi-2D structures with improved PL QY and stability.[132]

Colloidally stable ultrathin 2D halide perovskites are obtained by the solvent-assisted or sonication-assisted exfoliation of perovskite nanocrystals in the presence of excess surface ligands.[129–131] In solvent-assisted exfoliation, nanoplatelets of halide perovskites are formed by increasing the dilution of the colloidal nanocrystal solution. The increase in the osmotic swelling of nanocrystals caused by the solvent molecules at higher dilutions results in their fragmentation into nanoplatelets, and the excess ligand molecules which are already present in or subsequently added penetrate the interlayer region and stabilize the newly formed surface. Urban and co-workers demonstrated the solvent-assisted fragmentation of MAPbX$_3$ (X=Br, I) perovskite nanocrystals into stable nanoplatelets of different thicknesses by varying the degree of dilution or ligand concentration.[129] Such fragmentation of larger nanocrystals into quasi-2D structures is also induced by external stimuli such as sonication. Here, instead of dilution, sonication fragments the nanocrystals into thin perovskite layers, and the organic ligands penetrate these layers to bind with the newly formed facets.[130] The 2D perovskite nanoplatelets with different thicknesses are then separated by centrifugation at different speeds.

## 4. Optoelectronic Properties

In 2D halide perovskites, an interplay between defect passivation, charge-carrier confinement, and crystal structure and alignment leads to unique carrier dynamics and optoelectronic properties when compared to the 3D ABX$_3$ perovskites.[133,134] Photoexcitation of bulk crystals and films of 3D halide perovskites mainly generates free charge carriers by the thermal dissociation of excitons, and the carrier dynamics are dominated by the bimolecular electron-hole recombination.[16–18] This happens because the low exciton binding energy in these materials varies in a wide range (~2 to 60 meV)[24] which is comparable to or even below the thermal energy (~$k_B T$, where $k_B$ is Boltzmann constant) at room temperature. On the other hand, in layered 2D halide perovskites, the photogenerated charge carriers are strongly electronically confined along with the thickness of the inorganic layer. While the organic spacer cations passivate and reduce the contribution of defects towards the non-radiative monomolecular carrier recombination, contribution from excitonic radiative recombination dominates in the ultrathin 2D halide perovskites with high exciton binding energy,[133] which is irrespective of their lateral size. Further, in layered 2D halide perovskites, the increased Coulombic interactions due to the strong electronic confinement may enhance the Auger recombination which influences the bi-exciton and hot-carrier dynamics.[134] Nevertheless, in 3D ABX$_3$ perovskites, the excitonic monomolecular recombination dominates only in nanocrystals or quantum dots with high exciton binding energy when their size becomes comparable to or smaller than the corresponding exciton Bohr radius.[109,135–137]

*4.1 Dimensional and dielectric confinements*

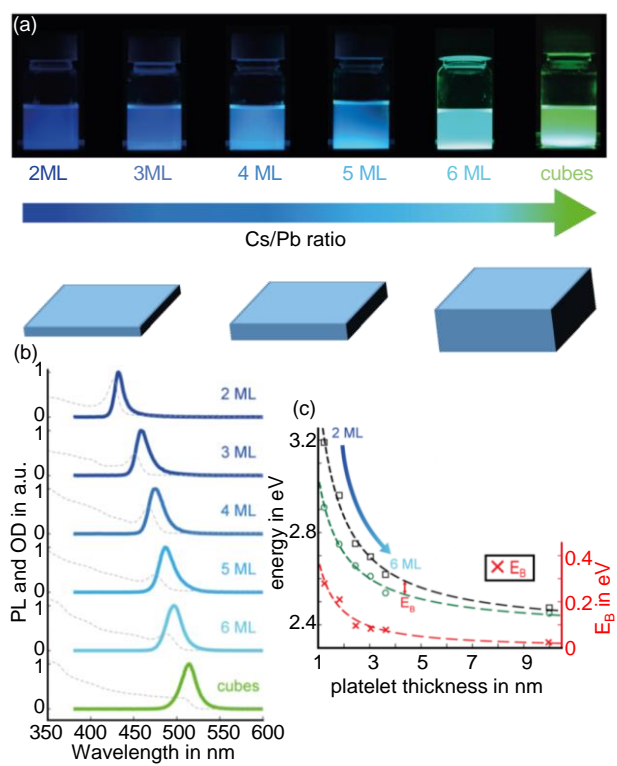

**Figure 7.** Dimensional confinement in colloidal perovskite nanoplatelets. (a) Photographs of CsPbBr$_3$ perovskite nanoplatelet solutions under UV light showing the thickness/shape-dependent color. (b) Absorption (dashed lines) and PL (solid lines) spectra of CsPbBr$_3$ nanoplatelets with different thicknesses. (c) The plots of exciton binding energy ($E_B$, red crosses) as the function of nanoplatelet thickness. $E_B$ is calculated as the difference between the continuum absorption onset energy $E_C$ (black squares) and the 1s exciton transition energy $E_{1s}$ (green circles) obtained from the corresponding absorption spectra. The dashed lines are for eye guidance. Reproduced with permission from ref. 116. Copyright (2018) American Chemical Society.

In bulk crystals and films of 3D ABX$_3$ perovskites, the optical bandgap and emission color can be tuned by changing the halide ion composition.$^{138–141}$ On the other hand, in dimensionally-confined halide perovskites, the change in shape from 3D nanocubes to 2D nanoplatelets and a decrease in the corresponding thickness blue shifts the bandgap and changes the emission color, as shown in Figure 7 a-c.$^{22,110,116}$ The excitons are dimensionally confined to the thickness of the perovskite nanoplatelets, and the exciton binding energy increases when the nanoplatelets become thinner. The excitonic confinement is characterized by a sharp excitonic peak and narrow-band emission (Figure 7b). In CsPbBr$_3$ perovskite nanoplatelets, the 1s exciton binding energy increases significantly up to 280 meV when the thickness is decreased to 2 monolayers (Figure 7c).$^{116}$

The dimensional and dielectric confinements of excitons in 2D halide perovskites are greatly enhanced when a dielectric mismatch between the perovskite layer and the surrounding organic cation spacer is involved.$^{45,60,61,123,142–145}$ As shown in Figure 8a, the inorganic BX$_6$ octahedral layers in layered 2D halide perovskites act as natural multi-quantum wells which are sandwiched between the barrier of adjacent organic spacer layers. Any photogenerated excitons are confined within the inorganic well, resulting in sharp excitonic absorption and narrow-band emission. Figure 8 e and f show the n-dependent variation of bandgaps in Pb- and Sn-based 2D R-P layered perovskites. The optical bandgap energy decreases as the number of inorganic layers increases in (BA)$_2$A$_{n-1}$B$_n$I$_{3n+1}$ [B=Sn, Pb] perovskites.$^{142}$ Interestingly, layered 2D halide perovskites containing MA cation in A-site showed larger bandgaps when compared to those with FA cation. This can be attributed to the octahedral tilt observed in MA-based halide perovskites due to its smaller size compared to the FA cation.$^{18,24}$ Nevertheless, such octahedral tilt is more prominent in layered 2D halide perovskites when the composition of the organic spacer cations is changed (Figure 3),$^{33,34,50,52,54}$ whose influence on the optoelectronic properties is discussed later. The exciton binding energy shows inverse scaling with the exciton reduced mass ($1/\mu=1/m_e+1/m_h$, where $m_e$ and $m_h$ are electron and hole masses, respectively) and the perovskite layer thickness more strongly in the dielectrically-confined 2D halide perovskites.[144] Table 2 lists the binding energy and reduced mass of excitons in some 3D and 2D halide perovskites at different temperatures. The optical transitions at room temperature can change from excitonic to free carriers when the thickness of the perovskite layer is too large such that n~∞.[158] Figure 8b shows the temperature dependence of exciton dissociation in (BA)$_2$(MA)$_{n-1}$Pb$_n$I$_{3n+1}$ perovskite with n=1 to 4. The fraction of free charge carrier is very low for n=1 over a wide temperature range, while it increases steeply with temperature for n≥2 and attains a plateau at a very high temperature.[145] This indicates that the high exciton binding energy in monolayered 2D halide perovskites favors the monomolecular recombination of bound excitons over the bimolecular recombination of free charge carriers.

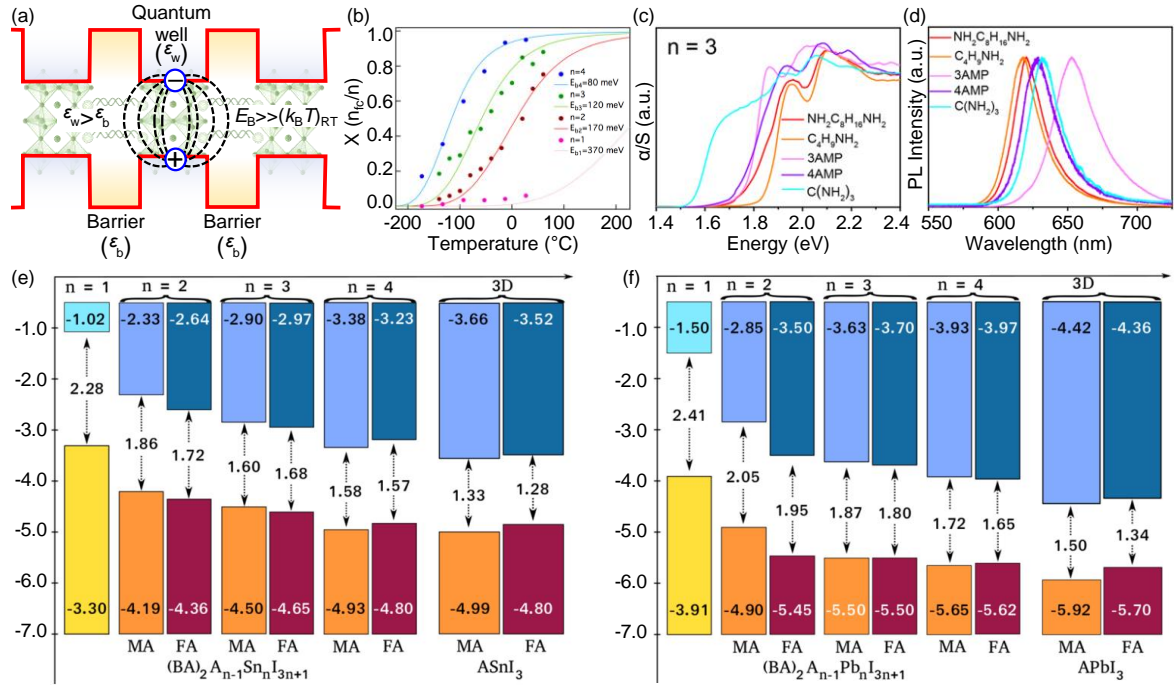

**Figure 8.** Optoelectronic properties of 2D R-P or D-J layered perovskites. (a) Scheme showing multi-quantum well structure of 2D layered perovskites. (b) Fraction of free charge carriers in $(BA)_2MA_{n-1}Pb_nI_{3n+1}$ as the function of temperature. The lines show simulated values using the Saha equation (at excitation density=$5\times10^{14}$ cm$^{-3}$), and the symbols indicate the experimental values. Reproduced with permission from ref. 145. Copyright (2017) American Chemical Society. (c) Absorption and (d) PL spectra of different 2D R-P and D-J perovskites of the same thickness (n=3) showing the effect of organic cation spacers on the optical properties. Reproduced with permission from ref. 53. Copyright (2018) American Chemical Society. (e,f) Bandgap energies of $(BA)_2A_{n-1}B_nX_{3n+1}$ perovskites with different compositions and thicknesses. Reproduced with permission from ref. 142. Copyright (2020) American Chemical Society.

**Table 2** Exciton binding energy ($E_B$) and exciton reduced mass ($\mu$) of 2D and 3D halide perovskites at different temperatures ($T$). Here, $m_0$ is the free electron mass, BDA is butane-1,4-diammonium, and DMPD is *N,N*-dimethylpropane-1,3-diammonium.

| | Dimensionality (D)/size | n | $E_B$ (meV) | $\mu$ | T (K) | Ref. |
|---|---|---|---|---|---|---|
| MAPbI$_3$ | 3D/bulk | ∞ | ≤12 | 0.104$m_0$ | 155−190, ~290, ~RT | 146–151 |
| | | | 16 | 0.104$m_0$ | 2 | 146–148 |
| MAPbBr$_3$ | 3D/bulk | ∞ | 14 | — | RT | 149 |
| | | | 25 | 0.117$m_0$ | 2 | 147 |
| CsPbBr$_3$ | 3D/bulk | ∞ | 33 | 0.126$m_0$ | RT | 152 |
| CsPbBr$_3$ | 2D/nanoplatelet | 2 | 280 | — | RT | 116 |
| (BA)$_2$MA$_{n-1}$Pb$_n$I$_{3n+1}$ | 2D/layered R-P | 1 | 467 | 0.221$m_0$ | 4 | 144 |
| | | 2 | 251 | 0.217$m_0$ | 4−290 | 144 |
| | | 3 | 177 | 0.201$m_0$ | 4−290 | 144 |
| | | 4 | 157 | 0.196$m_0$ | 4−290 | 144 |
| | | 5 | 125 | 0.186$m_0$ | 4−290 | 144 |
| (HA)$_2$MA$_{n-1}$Pb$_n$I$_{3n+1}$ | 2D/layered R-P | 1 | 361 | 0.18$m_0$ | 5 | 153,154 |
| | | 2 | 260 | — | 5 | 154 |
| | | 3 | 150 | — | 5 | 154 |
| | | 4 | 100 | — | 5 | 154 |
| (PEA)$_2$MA$_{n-1}$Pb$_n$I$_{3n+1}$ | 2D/layered R-P | 1 | 220 | — | 10 | 155 |
| | | 2 | 170 | — | 10 | 155 |
| | | 3 | 125 | — | 290 | 156 |
| | | 4 | 100 | — | 290 | 156 |
| BDAPbI$_4$ | 2D/layered D-J | 1 | 390 | — | RT | 157 |
| DMPDPbI$_4$ | 2D/layered D-J | 1 | 270 | — | RT | 157 |

Different compositions of the organic spacer cations induce structural variations in layered 2D halide perovskite crystals which change the optoelectronic properties of the material. For example, Figure 8 c and d show the absorption and PL spectra of 2D R-P and D-J layered perovskites with different compositions of organic spacer cations. The variation in the optical properties is correlated with Δ$d$, equatorial Pb-X-Pb bond angle distortion, and $\sigma$. For example, (C$_4$H$_9$NH$_3$)$^+$ cation induces greater Δ$d$ and $\sigma$ in the PbI$_6$ octahedron of an n=3 2D halide perovskite when compared to the (3AMP) cation.[75] As a result, the former shows an increase in the bandgap energy and blue-shifted emission. In other words, an increase in the octahedral distortion decreases the overlap between the metal and halide orbitals, which in turn decreases the band dispersion and increases the bandgap energy.[33,34,50,52,54] Such a change in the bandgap energy with the distortion of the inorganic framework has been widely studied in 3D ABX$_3$ analogs where high pressure is applied to compress the perovskite lattice.[159–163] Nevertheless, the modulation of the absorption and PL spectra and the exciton binding energy with the composition of organic spacer cation in 2D layered perovskites is strongly correlated to the change in dielectric properties of the material (Figure 9).

The exciton binding energies of 2D ($E_B^{2D}$) and 3D ($E_B^{3D}$) halide perovskites are related to the dielectric constants of the inorganic quantum well layer ($\varepsilon_w$) and the organic barrier layer ($\varepsilon_b$) by the following equation,[61,143]

$$E_B^{2D} = 4 \left[\frac{\varepsilon_w}{\varepsilon_b}\right]^2 E_B^{3D} \qquad 3.$$

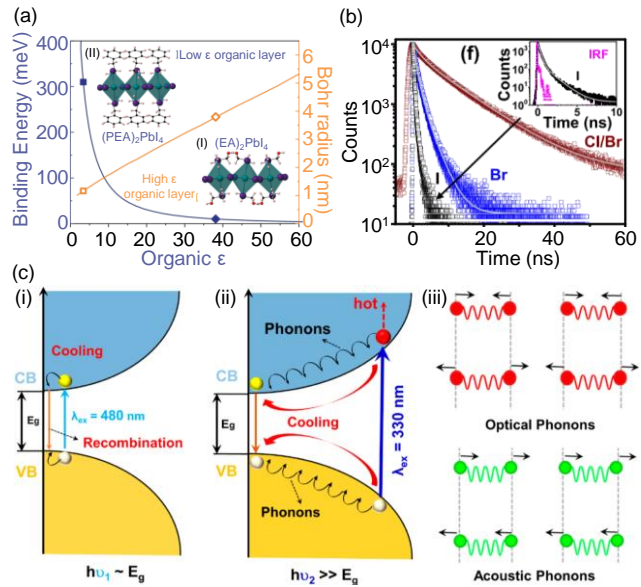

**Figure 9.** Dielectric confinement in layered 2D halide perovskites. (a) The binding energy and the Bohr radius of excitons as a function of organic barrier dielectric constant in (PEA)$_2$PbI$_4$ [square] and (EA)$_2$PbI$_4$ [diamond] perovskites. The inset shows the crystal structures of the 2D layered perovskites. Reproduced with permission from ref. 60. Copyright (2018) Nature Publishing Group. (b) PL decay profiles of (PEA)$_2$PbX$_4$ [X=Cl, Br, I] perovskite films. The arrow shows a faster PL decay from Cl/Br to Br and Br to I. The inset shows the PL decay profile of a (PEA)$_2$PbI$_4$ perovskite and the instrument response at a shorter time window. Reproduced with permission from ref. 61. Copyright (2020) American Chemical Society. (c) Carrier relaxation processes in layered 2D halide perovskites upon (i) band-edge excitation, (ii) hot-carrier generation, and (iii) phonon-assisted hot-carrier cooling. Reproduced with permission from ref. 164. Copyright (2019) American Chemical Society.

Thus, for a 2D halide perovskite with $\varepsilon_w > \varepsilon_b$, the Coulomb screening between the electrons and holes is less and the exciton binding energy is high. Figure 9a compares the calculated binding energy and Bohr radius of excitons in (PEA)$_2$PbI$_4$ and (EA)$_2$PbI$_4$ [EA=ethylammonium] perovskites with different dielectric constants of the organic spacer cation. In (PEA)$_2$PbI$_4$ perovskite the organic spacer PEA has a dielectric constant ($\varepsilon_b$~3.3) smaller than that of the inorganic well ($\varepsilon_w$~6.1).[60] Therefore, the exciton Bohr radius decreases (~1 nm), and the exciton binding energy increases (>300 meV), showing strong dielectric confinement. Also, the strongly bound exciton undergoes fast radiative recombination which increases the PL QY of the dielectrically-confined 2D halide perovskites.[143] On the other hand, in (EA)$_2$PbI$_4$ perovskite the dielectric constant of EA ($\varepsilon_b$~37) is more than 10 times greater than that of PEA, which is in turn much greater than $\varepsilon_w$. This weakens the dielectric confinement and enhances the Coulomb screening between the electron and the hole. As a result, the exciton Bohr radius increases (~4 nm), and the exciton binding energy decreases (<15 meV).[60] In addition to the highly polarizable organic spacer cations, the electronic confinement in 2D halide perovskite nanoplatelets and R-P layered structures is weakened by replacing or combining the organic spacer cations with the electron-accepting organic chromophores such as perylene diimide and the organic charge-transfer complexes such as pyrene alkylammonium-tetracyanoquinodimethane (pyrene-C4:TCNQ) donor-acceptor system.[165,166] For example, photoexcitation of a 2D (pyrene-C4:TCNQ)$_2$PbI$_4$ perovskite resulted in the long-lived (1-4 µs) free charge carriers in the inorganic layers which was accredited to charge-separated holes transferred from the (pyrene-C$_4$:TCNQ) charge-transfer complex into the inorganic perovskite layers. On the other hand, the electrons were confined in the TCNQ molecules. Such strategies are useful in achieving the charge-separated states in electronically-confined 2D halide perovskites for their applications to solar cells, photodetectors, and photocatalysis. Further, Nag and coworkers have shown that the dielectric confinement in layered 2D halide perovskites can also be tuned by changing the composition of halide ions.[61] Change in the halide composition from Cl to Br and Br to I increases $\varepsilon_w$ of the inorganic (quantum well) layer in (PEA)$_2$PbX$_4$ perovskites which increases the dielectric confinement of excitons. As a result, the fast excitonic recombination decreases the PL lifetime of (PEA)$_2$PbX$_4$ perovskite while changing the composition from Cl to Br and Br to I (Figure 9b). This is a trend observed opposite to the 3D ABX$_3$ perovskites where the PL lifetime increases from Cl to Br and Br to I.[22,109]

Organic spacer cation also influences the hot-carrier dynamics in layered 2D halide perovskites.[134,164,167–170] Photoexcitation of 2D halide perovskites at band edge generates strongly bound excitons which can undergo monomolecular recombination to produce photons [Figure 9c (i)]. Photoexcitation above bandgap generates hot electrons and hot holes which occupy the higher electronic states. These hot carriers must thermalize non-radiatively to the band edge by emitting phonons before they undergo radiative recombination [Figure 9c (ii)]. The released phonons can interact with the hot-carrier relaxation [Figure 9c (iii)] and affect the fast-carrier dynamics. An increase in the dielectric constant of the organic spacer cation in layered 2D halide perovskites increases the hot-carrier relaxation time from hundreds of femtoseconds to picoseconds.[164,167,170] Further, the slow propagation of acoustic phonons and their upconversion to optical phonons in 2D halide perovskites by varying the composition of organic spacer cation and increasing the number of inorganic layers (n) can heat the cold carriers and further delay the hot-carrier cooling to thousands of picoseconds.[171] Hot-carrier cooling in such 2D halide perovskites is much longer than in the case of 3D ABX$_3$ perovskites.[168] The prolonged hot-carrier relaxation in 2D layered perovskites provides means to harvest hot electrons and holes for optoelectronic and photocatalytic applications.

*4.2 Self-trapped excitons and broad photoluminescence*

The optoelectronic properties of halide perovskites are greatly influenced by temperature. At higher temperatures, the electron-phonon interaction broadens the absorption and PL bands inhomogeneously in 3D ABX$_3$ perovskites.[172–174] On the other hand, the sharpening of absorption and PL peaks is observed at cryogenic temperature, most likely due to the low electron-phonon interaction. In addition to the spectral shape, the position of the absorption and emission peak also shifts with temperature which is attributed to the change in octahedral distortion, tilt angles, or crystal phase.[78,175] In layered 2D halide perovskites, the influence of temperature on the optoelectronic properties is rather unique. These layered 2D structures show largely Stokes-shifted and intense broadband emission at lower temperatures, depending on the type of organic spacer cations.[176–178] In some of the layered 2D halide perovskites and low-dimensional post-perovskite chains, broad emission (BE) covering most of the visible spectrum has been observed even at room temperature, which makes them promising white light-emitting fluorophores.[176,179–181] The broadband emission from layered 2D halide perovskites is known to originate from the self-trapped excitons (STEs).

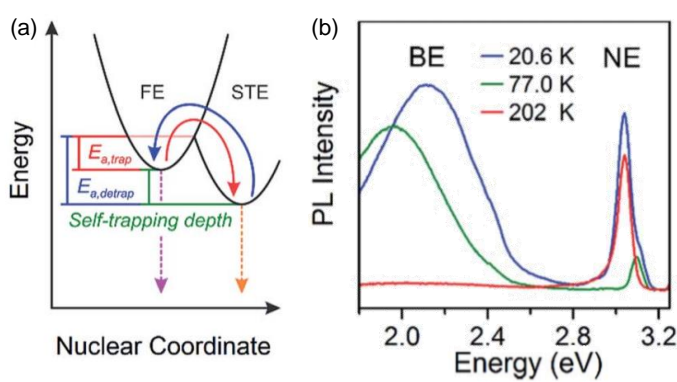

**Figure 10.** Exciton self-trapping in layered 2D halide perovskites. (a) Scheme showing the exciton self-trapping (red arrow) and detrapping (blue arrow) with free exciton (FE) emission in pink and STE emission in orange ($E_{a,trap}$=activation energy for self-trapping and $E_{a,detrap}$=activation energy for detrapping). (b) PL spectra of (HIS)PbBr$_4$ (HIS=histammonium) perovskite single crystal at different temperatures showing the narrow emission (NE) at ca. 3 eV and the broad emission (BE) at ca. 2 eV. Reproduced with permission from ref. 176. Copyright (2017) Royal Society of Chemistry.

STEs are the photoexcited bound electron-hole pairs that are trapped and stabilized into a self-created potential well.[177] The electron (hole) forms a small localized polaron by the virtue of a deformable lattice which then attracts the hole (electron). The BE from layered 2D halide perovskites is observed at low temperatures at which the formation of STEs is favored. At room temperature, on the other hand, detrapping of STEs results in the narrow emission (NE) with a small Stokes shift from the radiative recombination of the free exciton (FE).[182] Here, FEs are those which are not self-trapped, and they should not be confused with the unbound excitons or free charge carriers. Figure 10a shows the exciton self-trapping and detrapping in a layered 2D halide perovskite. Karunadasa and coworkers have shown the formation of STEs and broadband emission from a variety of layered 2D halide perovskites and similar materials.[176,180,183,184] Figure 10b shows the temperature-dependent PL spectra of (HIS)PbBr$_4$ perovskites which show NE at 202 K and BE at or below 77 K. The NE band can be observed even at cryogenic temperature along with the BE band which is due to the thermal equilibrium between the FEs and STEs. The BE in layered 2D halide perovskites is correlated to the structural distortions induced by different organic spacer cations.[177,182] Unlike the NE from FEs at room temperature which is influenced by the in-plane Pb-X-Pb distortion (Figure 8 c and d), the BE in layered 2D halide perovskites from STEs is affected by the out-of-plane distortion. For example, (BA)$_2$PbBr$_4$ perovskite with least out-of-plane distortion ($\theta_{out}$=177°) shows only NE at the wide temperature range from 200 to 20 K, while (HIS)PbBr$_4$ with relatively higher out-of-plane distortion ($\theta_{out}$~157°) shows BE at lower temperatures (<200 K).[176] Another example is highly distorted (AEA)PbBr$_4$ perovskite ($\theta_{out}$~22°) which shows dominant BE at room temperature.[176] Apart from the organic spacer cation-induced structural distortion, halogens also influence the self-trapping of excitons and BE in layered 2D halide perovskites.[178,181] Gautier et al. showed that STEs are more favored than FEs for Cl-based layered 2D halide perovskites than for Br- or I-based ones.[181] This is attributed to the lower self-trapping depth in the case of Br and I when compared to Cl. In other words, the detrapping barrier is high for lighter halides. Halogen composition influences lattice deformation through different metal-halide bonding, which in turn influences the trapping/detrapping of excitons in layered 2D halide perovskites.

STEs are considered as intrinsic as the trapping potential is originated from the excitons themselves. However, there exists a debate among perovskite researchers that whether the origin of broadband emission is intrinsic, or is caused by the factors such as dopants and defects.[185–187] Recently, Kahmann et al. showed the extrinsic nature of BE in PEA- and its fluorinated derivative (FPEA)-based 2D R-P layered perovskites.[185] While the STEs cannot be accessed directly from the ground state, their work showed the existence of in-gap (defect) states accessed by the below bandgap excitation which contributes to the broadband emission. Further, the sample-to-sample variation in the intensity of NE and BE substantiated the presence of emissive in-gap states with different densities in different samples. Also, Yin et al. showed that the Stokes-shifted and broadband emission from $(PEA)_2PbI_4$ perovskites is originated from the defect states.[186] These radiative defect states are contributed by the surface or bulk iodine vacancies, and Yin et al. demonstrated the suppression of such broadband emission by passivating the defects using excess PEAI.

## 5. Environmental Stability

Despite the outstanding optoelectronic and charge carrier properties of 3D $ABX_3$ perovskites that are promising for high-efficiency light-emitting and light-harvesting devices, these materials often suffer from poor stability against moisture and light.[18] Various studies have revealed the formation of mono- or dihydrate species as the first step towards the degradation of halide perovskites when they are exposed to moisture or water vapor (Figure 11 a).[28,188–197] The hydrated 3D $ABX_3$ perovskite decomposes irreversibly in the presence of excess water into aqueous HX and solid $BX_2$, and the organic A-site cation is either released as gas or dissolved in water. The presence of halide vacancies, surface defects, and surface reactivity in halide perovskites accelerate the decomposition process by acting as active sites for adsorbing water and oxygen molecules.[28,29,191,198,199]

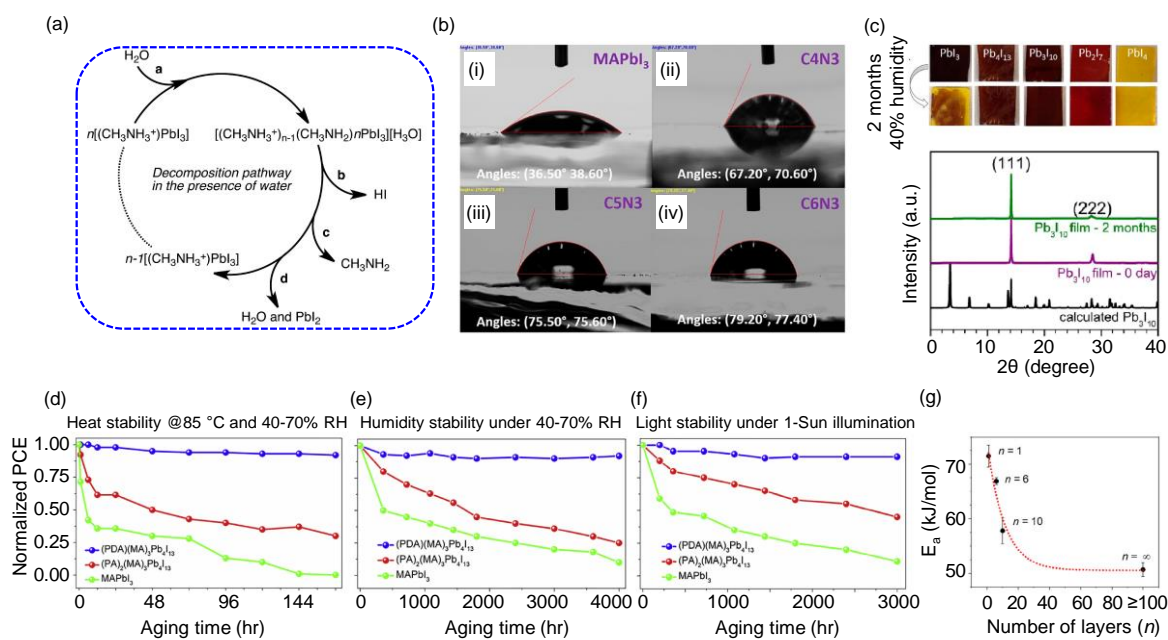

**Figure 11.** Environmental stability of halide perovskites. (a) Scheme showing the hydration and decomposition cycle of MAPbI$_3$ perovskite in the presence of water/moisture. Reproduced with permission from ref. 188. Copyright (2014) American Chemical Society. (b) Contact angles between a water droplet and the perovskite films (i) MAPbI$_3$, (ii) (BA)$_2$MA$_2$Pb$_3$I$_{10}$, (iii) (PA)$_2$MA$_2$Pb$_3$I$_{10}$, and (iv) (HA)$_2$MA$_2$Pb$_3$I$_{10}$. Reproduced with permission from ref. 50. Copyright (2019) American Chemical Society. (c) Photographs and XRD of films of halide perovskites with 3D and layered 2D structures that are exposed to humidity for over 2 months. Reproduced with permission from ref. 77. Copyright (2015) American Chemical Society. Comparison of (d) heat-, (e) moisture-, and (f) photostability of solar cells constructed from 3D and layered 2D halide perovskites. The labels of the y-axis in (e) and (f) correspond to that of (d). Reproduced with permission from ref. 72. Copyright (2018) Elsevier Inc. (g) The plot of activation energy ($E_a$) of halide exchange as the function of inorganic layer thickness (n) of a layered 2D halide perovskite. Reproduced with permission from ref. 48. Copyright (2020) American Chemical Society.

Halide vacancies in 3D $ABX_3$ perovskites are very common due to their low energy of formation.[28,198,200,201] Furthermore, these perovskites exposed to ambient air generate superoxide under photoirradiation which induces oxidative degradation of the material.[27,29,198] For example, photogenerated electrons in $MAPbI_3$ perovskite nanocrystals or films reduce the oxygen molecules adsorbed on the vacant iodide sites to superoxide.[27,29] The thus generated superoxide degrades the material by reacting with the photo-oxidized perovskite structure to form methylamine, $PbI_2$, iodine, and water. In addition to the moisture- and superoxide-induced degradation, migration of halide ions under prolonged photoirradiation or electrical bias is another detrimental effect that limits the operational stabilities and efficiencies of 3D $ABX_3$ perovskite-based devices.[26,202–208] The migration of halide ions in the perovskite lattice is a thermally activated process that is assisted through the halide vacancy defects. It has the lowest activation energy (~0.58 eV for iodide ion migration) when compared to the migration of other ions in the perovskite lattice.[202] Such easy movement of halide ions under light irradiation or the influence of an electric field during device operation results in defect migration, phase segregation, or lattice distortion inside the perovskite layers which ultimately destabilizes the material if such processes are prolonged and irreversible. Further, the halide ion migration across the halide perovskite-electrode interface leads to the degradation of the perovskite structure and electrodes through the extraction of ions from the perovskite layer.[26,209,210] 2D R-P or D-J layered perovskites, on the other hand, demonstrate improved stability against moisture and light (Figure 11 b-g).[48,50,59,72,77,211] Nevertheless, long-term stability of colloidal 2D perovskite nanoplatelets/nanosheets is still a challenge.[123,212] In this section, we discuss the stability of 2D halide perovskites concerning moisture, heat, and light.

*5.1 Moisture and thermal stability*

In layered 2D halide perovskites, the moisture stability is rendered by the hydrophobic nature of aliphatic or aromatic organic ammonium spacer cations.[48,50,53,59,72,77,211,213–215] Such hydrophobic nature of layered 2D halide perovskites is characterized by a high average contact angle of the water droplet on the perovskite film. As shown in Figure 11 b, the contact angle is lowest for the MAPbI$_3$ perovskite film, while it increases with an increase in the organic chain length of the spacer cation in R$_2$MA$_{n-1}$Pb$_n$I$_{3n+1}$ perovskites, indicating an increase in the hydrophobicity of the material.[50] The layered 2D halide perovskite crystals with hydrophobic organic spacer cations can orient themselves on the substrates and passivate the grain boundaries, forming densely packed films. This decreases the likelihood of direct moisture contact, water penetration, or oxygen diffusion, and accordingly prevents the first step degradation by preventing the formation of hydrated perovskite species. For instance, the films of (BA)$_2$(MA)$_2$Pb$_3$I$_{10}$ perovskites are stable for over two months maintaining their crystal structure under high humidity, while the MAPbI$_3$ perovskite film decomposes to yellow PbI$_2$ in a short time (Figure 11 c).[77] Karunadasa and coworkers reported the formation of highly oriented and densely packed (PEA)$_2$(MA)$_2$Pb$_3$I$_{10}$ perovskite films which show enhanced moisture stability compared to the MAPbI$_3$ when exposed to a 52% relative humidity (RH) over 46 days.[59] Tsai et al. used the hot-casted films of 2D R-P layered perovskites to construct a solar cell and encapsulated it with epoxy which exhibited prolonged operational stability (ca. 94 days) in the presence of light and high humidity (RH~65%).[213] Ahmad et al. showed that 2D D-J layered perovskites are more stable than 2D R-P layered or 3D ABX$_3$ perovskites by demonstrating a much-improved performance of the unencapsulated solar cell device under heat (Figure 11 d), moisture (Figure 11 e), and light exposure (Figure 11 f).[72] Cho et al. showed that the use of fluorinated organic spacer cation (CF$_3$)$_3$CO(CH$_2$)$_3$NH$_3$$^+$ increases the hydrophobicity of the layered 2D halide perovskites, making them water repellent.[214] Such hydrophobic 2D halide perovskites can be blended with or deposited on 3D ABX$_3$ perovskites to obtain a 2D/3D composite that is moisture and oxidation proof.[214,216]

Better thermal stability in layered 2D halide perovskites when compared to 3D ones is attributed to the stronger electrostatic interaction between the organic spacer cation and the $BX_6$ octahedra and the van der Waals interaction among the stacked inorganic layers. In 2D R-P layered perovskites, the monoammonium organic spacer cations create the van der Waals gap between the adjacent stacking layers. On the other hand, the gap is completely removed in the case of 2D D-J layered perovskites where the diammonium organic spacers link the adjacent layers electrostatically. Hence, the 2D D-J layered perovskites become more stable towards heat than their R-P counterparts.[72] The thermal stability of the layered 2D halide perovskites also is affected by the orientation of crystals on the substrate. The vertically oriented 2D halide perovskites are thermally more stable than the parallelly or randomly oriented ones which are attributed to the better heat distribution among the vertically oriented crystallites in the film.[50,53] Nevertheless, the thermal and the air stability of the layered 2D halide perovskite films also depends on the deposition method, the film thickness, and the substrate crystallinity.[50]

*5.2 Photostability*

The bulky organic spacer cations enhance the photostability of layered 2D halide perovskites (Figure 11 f).[50,72,213] The efficient passivation of perovskite lattice against moisture and oxygen by the hydrophobic organic spacers reduces the formation of reactive oxygen species in the presence of light which improves the photostability.[217] More importantly, the photostability is improved by the suppression of the halide ion migration in layered 2D halide perovskites.[48,211,218,219] The out-of-plane migration of halide ions in 2D halide perovskites, which is from one inorganic slab to another stacking together, is prevented by the bulky organic spacer cations that act as the barrier layer in between. On the other hand, the in-plane movement of halide ions within the perovskite layer is suppressed by the low probability of point defect formation due to the large formation energy of these point defects. For example, $(BA)_2(MA)_2Pb_3I_{10}$ perovskite has a halide vacancy formation energy of ~5.46 eV, which is higher than that of MAPbI$_3$ perovskite (3.44 eV).[211] Therefore, the vacancy-mediated diffusion of halide ions in layered 2D halide perovskites is suppressed. Further, the activation energy for the halide exchange increases with decreasing the number of inorganic layers n<10 in layered 2D halide perovskites (Figure 11 g).[48] This indicates the increased thermal barrier for the halide ion migration in layered 2D perovskite structures when compared to their 3D analogs.

While layered 2D halide perovskites are relatively more stable than the 3D ABX$_3$, the stability of these low dimensional structures in their colloidal states is still a question, especially during long-time room temperature storage and after post-synthetic purification.[36,74,123,212,220] Multiple washing steps after synthesis can strip off the ligands from the surface of the perovskite nanoplatelets or nanosheets and thus destroy the colloidal stability.[36] Removal of a large number of ligands can also lead to low PL QY and material degradation, owing to the formation of defects.[116] Moreover, the use of too polar antisolvent during the purification step can dissolve the perovskite nanoplatelets.[212] Therefore, it is plausible to implement efficient strategies to stabilize colloidal 2D halide perovskites. The addition of surface ligands and halides to the washing solvent during the purification steps or to the colloidal solution after purification is one way to preserve the ligand density on the surface of perovskite nanoplatelets or nanosheets. This also helps to minimize the surface defects and improve the PL QY. Also, washing with solvents that have low dielectric constants such as diethylene glycol dimethyl ether (diglyme) has shown to be efficient for maintaining stable and highly luminescent colloidal halide perovskites.[221]

## 6. Applications

### 6.1 Solar cells

Since Miyasaka and co-workers first demonstrated the 3D ABX$_3$ perovskite solar cell in 2009 with an efficiency of 3.8%,[222] extensive study has been performed in the last decade to improve the open circuit potential ($V_{oc}$) and achieve high photon conversion efficiency (PCE).[1,223–225] According to the National Renewable Energy Lab (NREL) best research cell efficiency chart 2021, the current record-breaking PCE of halide perovskite solar cell is 25.5%,[226] which can be attributed not only to their intrinsic excellent optoelectronic properties but also to better material deposition strategies,[227] defect passivation,[228,229] and exploration of better electron or hole transport layers.[2] Despite a great potential for solar energy harvesting, 3D $ABX_3$ perovskites are not yet commercially viable due to their poor stability.[230] 2D halide perovskites such as R-P or D-J layered structures, on the other hand, are more stable alternatives to $ABX_3$ in solar cells. In this section, we discuss the developments in 2D halide perovskite-based solar cells.

In 2014, the 2D $(PEA)_2(MA)_2Pb_3I_{10}$ perovskite was used as the light-absorbing layer for the first time in a planer solar cell device which showed a PCE of 4.74 %.[59] Later, various 2D R-P and D-J layered perovskite-based solar cells were constructed with significant improvement in their efficiency.[34,56,64,69,70,72,73,77,126,213,215,219,231] Tsai et al. demonstrated a PCE of 12.51% with $V_{oc}$ of 1.01 V in $(BA)_2(MA)_3Pb_4I_{13}$-based solar cell.[213] The device was hysteresis-free in the current ($J$)-voltage ($V$) curve, which is essential for reproducible higher efficiencies. Such hysteresis in the $J$-$V$ curve is reported in the case of 3D $ABX_3$ perovskite solar cells, which is mainly attributed to the ionic conductivity and poor photostability of the material.[202,232,233] The quality and packing of layered 2D halide perovskite crystals in a film are influenced by the choice of organic spacer cations. This in turn affects the stability and efficiency of perovskite solar cells. Ahmad et al. reported a D-J layered structure-based perovskite solar cell with a PCE of 13.3%, which was higher than for R-P layered structure-based devices.[72] The improved performance of D-J layered structure-based perovskite solar cells was attributed to the formation of high-quality crystals with less octahedral distortion and dense packing which lead to better carrier mobility. Zheng et al. used benzylammonium, N,N-dimethylammonium, propyl-1,3-diammonium, and butyl-1,4-diammonium cations in layered 2D halide perovskites and constructed solar cells.[215] Among these, 2D halide perovskite films with benzylammonium as cations provided a high PCE (17.4%) and excellent operational stability (>500 h) under humid conditions (80% relative humidity). Also, an increase in the number of inorganic layers or thickness of the 2D halide perovskite increases the PCE in solar cells, which is possibly due to the small lattice distortion, low exciton binding energy, and long carrier diffusion in thick quasi-2D halide perovskite layers.[73] Nevertheless, the organic spacer cations act as an insulating layer which reduces the carrier diffusion between the stacked inorganic layers and suppresses the carrier extraction to the charge transport layers. Therefore, the orientation of layered 2D halide perovskite crystals on a substrate greatly determines the efficiency of solar cells.[77,107,213,234–236] The 2D halide perovskites can align in-plane (parallel/horizontal) or out-of-plane (perpendicular/vertical) to the substrate during their growth or deposition. Consequently, solar cells with vertically oriented layered 2D halide perovskites show better charge mobility and extraction and therefore high PCE (Figure 12). Hot-casting of precursors on the substrate results in the vertical growth of 2D halide perovskites (Figure 12 a) and better solar cell performance (Figure 12e) when compared to the room temperature or moderate temperature deposition, which forms parallelly or randomly oriented crystals (Figure 12 b).[107,213] Such preferred vertical orientation of perovskite crystals can be characterized by the concentrated spots in Grazing incidence wide-angle X-ray scattering (GIWAXS) patterns which smear out into rings for the randomly oriented ones (Figure 12 c and d). The efficient charge extraction from the strongly quantum confined 2D halide perovskite layers can also be achieved by attaching electron-accepting molecules such as fullerenes and organic chromophores on the perovskite surface.[165,237] Intercalation of inorganic moieties such as halogens within the van der Waals gap of organic spacer cation layers can also reduce the dielectric confinement in 2D halide perovskite layers.[238] This makes the excitons more viable for the dissociation and extraction to the charge transport layers in the solar cells.

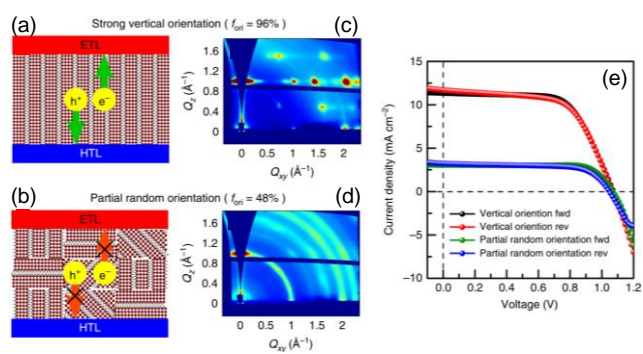

**Figure 12.** Performance of a layered 2D halide perovskite-based solar cell. (a,b) Scheme showing the (a) vertical and (b) random orientation of layered 2D halide perovskite crystals and the corresponding influence on the charge carrier extraction in a solar cell. (c,d) GIWAXS patterns of layered 2D halide perovskite films showing (c) vertical and (d) random orientation. (e) Current density versus voltage curves of layered 2D halide perovskite-based solar cells with vertical and random orientation of crystals. Reprinted with permission from ref. 107. Copyright (2018) Nature Publishing Group.

Besides, solar cells with 2D/3D perovskite heterostructures as light absorbers are gaining recent attention due to their extreme stability and high efficiency (PCE>20%).[214,239–243] In such perovskite solar cells, higher stability against moisture is due to the hydrophobicity of layered 2D halide perovskites, and high efficiency is achieved by the combined 2D/3D perovskite layer. The hydrophobic 2D perovskite layers also passivate the surface defects, reduce the ion migration, and suppress the nonradiative recombination in 3D $ABX_3$ perovskites. Zhu et al. demonstrated the PCE of 20% for the 2D/3D perovskite solar cells by incorporating 2D $(PEA)_2PbI_4$ perovskite nanosheets between 3D $MAPbI_3$ perovskite and hole-transporting layer.[243] The favorable band alignment between the graded 2D/3D perovskite structure and the hole-transport layer at the interface increased the hole transfer efficiency and reduced the interface charge recombination, which resulted in high PCE. Liu et al. reported a 2D/3D perovskite solar cell with excellent operational stability under ambient atmosphere and PCE >22% by depositing an (FEA)$_2$PbI$_4$ [FEA= pentafluorophenylethylammonium] layer on top of a 3D halide perovskite.[239] They demonstrated the enhancement of interfacial hole extraction in the device by the 2D perovskite layers. Bouduban et al. showed PCE >23% and very high $V_{oc}$ (>1.0 V) in solar cells with 2D/3D perovskite heterostructures which contain PEA or fluorinated derivatives of PEA as spacer cations.[242] They attributed the excellent solar cell performance to the planar orientation of layered 2D halide perovskites on top of 3D ones which not only passivate the defects but also prevent the back-electron transfer without affecting the charge extraction from the perovskite layer.

*6.2   LEDs and lasers*

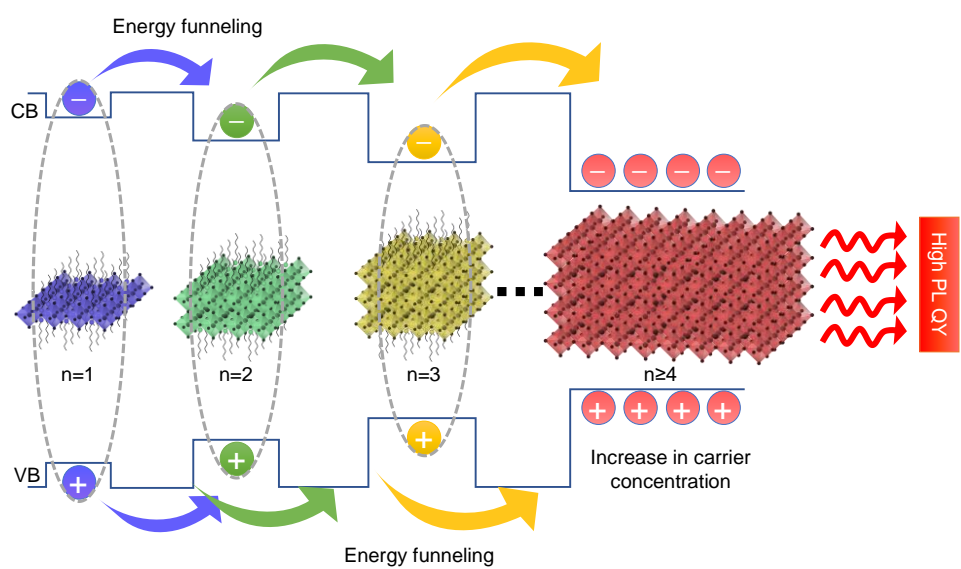

**Figure 13.** Scheme showing efficient energy transfer in quasi-2D halide perovskites. CB and VB are conduction and valence bands, respectively.

Strong electronic confinement in layered 2D halide perovskites leads to high PL QY and narrow-band emission which are promising for LED and lasing applications.[244–246] While LEDs based on 3D ABX$_3$ perovskites have already achieved an external quantum efficiency (EQE) of >20% but with limited stability,[18] 2D halide perovskite light-emitting devices are following the legacy with better stability. For example, green- and NIR-emitting quasi-2D perovskite LEDs have shown champion EQEs of >20%,[247–249] while >13% is reported for the red-emitting[250] and ~12% for the blue-emitting devices.[251] Here, we review the advancement of layered 2D halide perovskite-based LEDs and lasers.

Efficient energy funneling plays a vital role in improving the PL QY and overall performance of the layered quasi-2D halide perovskites in LEDs, low-threshold amplified spontaneous emission (ASE), and lasing.[245,246,252–256] In quasi-2D halide perovskites, the mixture of layers with different n creates an inhomogeneous distribution of energy states. As a result, energy funneling can take place from larger bandgap (smaller n) to smaller bandgap (larger n) domains (Figure 13). Efficient energy funneling increases the carrier concentration at the small bandgap 2D perovskite layers. Such high-density charge carriers recombine radiatively, and the PL QY of the material is increased even at low excitation intensities.[257] Figure 14 a shows the structure of a typical perovskite LED, where a quasi-2D halide perovskite nanosheet film is interposed as a light-emitting layer in between the hole transport [poly(3,4-ethylenedioxythiophene): polystyrene sulfonate, PEDOT: PSS] and electron transport [2,2′,2-(1,3,5-Benzinetriyl)-tris(1-phenyl-1-$H$-benzimidazole, TPBi] layers.[244] Electrodes such as indium tin oxide (ITO) and Al are attached to these charge transport layers for the electrical contact. Efficient energy funneling increases the efficiency of quasi-2D halide perovskite LEDs even at low carrier injections. Yuan et al. demonstrated energy funneling enhanced EQE of 8.8% at a low turn-on voltage of 3.8 V in a NIR LED by using n=5 quasi-2D $(PEA)_2(MA)_{n-1}Pb_nI_{3n+1}$ perovskite as a light-emitting layer.[258] The thicker quasi-2D halide perovskite film provided efficient energy transfer with better surface coverage. Multicolored quasi-2D halide perovskite LEDs are realized by tuning the thickness or the halide composition.[244,259–261] Figure 14 b shows the EL spectra of the quasi-2D halide perovskite nanosheet-based LEDs spanning a large visible spectrum from blue to red, and Figure 14 c shows the real multicolored quasi-2D halide perovskite nanosheet-based LED devices.

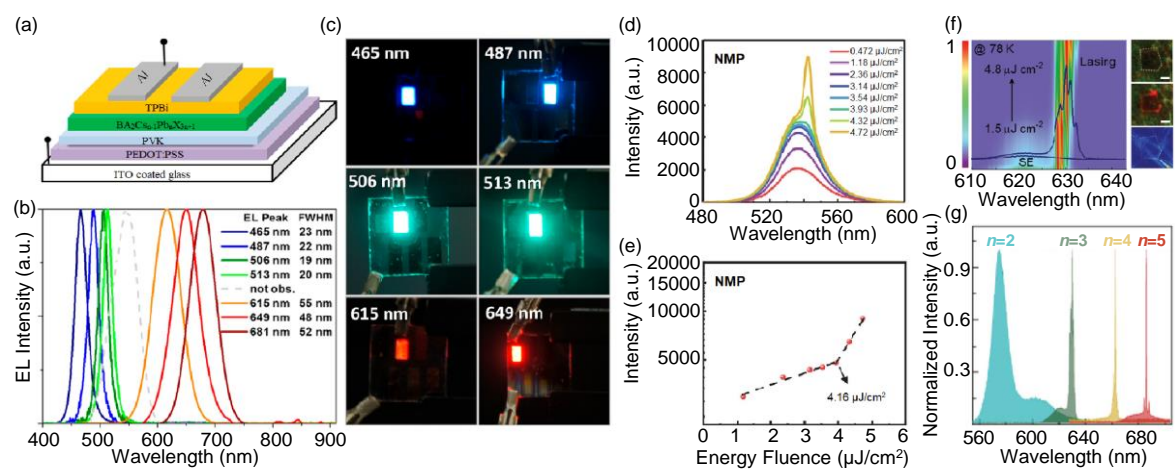

**Figure 14.** Light-emitting applications of layered 2D halide perovskites. (a-c) Multicolor LEDs based on quasi-2D $(BA)_2(Cs)_{n-1}Pb_n(Br/X)_{3n+1}$ [X = Cl, I] perovskite nanosheets: (a) LED device architecture, (b) electroluminescence spectra, and (c) real LED devices showing blue, green and red electroluminescence. Reprinted with permission from ref. 244. Copyright (2019) American Chemical Society. (d) ASE spectra and (e) ASE threshold as the function of pump fluence in quasi-2D $(PEA)_2(FA)_3Pb_4Br_{13}$ perovskite films. Reprinted with permission from ref. 245. Copyright (2020) Wiley-VCH. (f,g) Lasing from mechanically exfoliated $(BA)_2(MA)_{n-1}Pb_nI_{3n+1}$ perovskite microflakes: (f) Multimode lasing spectra as the function of pump fluence for n=3 2D layered perovskites and (g) multicolor lasing from the layered 2D halide perovskites with different n values. Reprinted with permission from ref. 246. Copyright (2019) Wiley-VCH.

The EQE of a quasi-2D halide perovskite LED is affected by the nonradiative recombination centers or defects, which are originated from the uncoordinated $Pb^{2+}$ ions in the lead-based perovskites,[247] oxidation of $Sn^{2+}$ to $Sn^{4+}$ in the tin-based perovskites,[262] grain boundaries, and ion vacancies.[263] Kong et al. showed a champion EQE of 20.5% and a luminance of 13,400 cd m$^{-2}$ in a green-emitting LED by passivating the defects in quasi-2D (BA)$_2$Cs$_{n-1}$Pb$_n$Br$_{3n+1}$ [n≥4] perovskite film using methylsulfonate.[247] Methylsulfonate also enhanced the energy funneling by increasing the donor-to-acceptor ratio through the redistribution of inorganic layer thickness. The strong exciton binding energy in the quasi-2D halide perovskites can induce non-radiative Auger processes which decrease the EQE of LED devices.[248,264] Recently, Jiang et al. reported a very high EQE of 20.36% and a champion luminance of 82,480 cd m$^{-2}$ in a quasi-2D halide perovskite LED by employing *p*-fluoro-PEA as the spacer cation, which was attributed to the lowing of the Auger recombination rate due to the reduced exciton binding energy.[248] While the decreased exciton binding energy lowers the PL QY by increasing the trap-assisted nonradiative recombination, compensation was made by passivating such defects using potassium trifluoromethanesulfonate. Nonetheless, the nonradiative recombination is also induced by the higher concentration of n=1 layers in the quasi-2D halide perovskites, which is due to either the presence of deep defect states in such wide-bandgap domains or the strong exciton-phonon coupling.[265–267] Therefore, it is essential to control the concentration and distribution of layers with different thickness in quasi-2D halide perovskites for high-efficiency LEDs. The formation of the n=1 layers can be controlled by modulating the interaction between the organic spacer cations and the inorganic layers and altering the crystallization process.[266–268] Alongside, optimally oriented and well-connected 2D perovskite domains render efficient charge transport.[265] Organic spacer cations can affect the performance of quasi-2D halide perovskite LEDs by tuning the organic barrier and quantum well width, modifying the carrier recombination pathways, influencing the energy funneling process, and affecting the film quality.[250,269,270] In fact, the efficiency of energy funneling in the quasi-2D halide perovskites depends upon the size/length of the organic spacer cations where the shorter chain spacer results in more efficient energy transfer.[36,58] Klein et al. showed efficient energy funneling in 2D MAPbBr perovskite nanosheets which were prepared by using short-chain dodecylamine ligands.[36] This resulted in red-shifted emission with an increase in the PLQY up to 49% for the sample with n=4. On the other hand, 2D halide perovskite nanosheets synthesized using relatively long-chain ligands such as tetradecylamine and hexadecylamine showed weaker energy funneling characterized by the emission at higher energy and with lower quantum yield.

An increase in the local exciton density through efficient energy funneling in the quasi-2D halide perovskites can result in population inversion, which is a prerequisite for ASE or lasing. Lei et al. demonstrated that efficient energy funneling leads to ASE from films of quasi-2D $(PEA)_2(FA)_3Pb_4Br_{13}$ perovskite at a low threshold of 4.16 µJ cm$^{-2}$ (Figure 14 d and e).[245] They also demonstrated distributed feedback lasing by optically pumping the quasi-2D halide perovskite film at room temperature using a pulsed laser. The distributed feedback cavity was constructed by nanoimprinting the 1D grating with a period of 350 nm on a substrate and spin coating the 2D halide perovskites on the top of it. The lasing threshold was 10 µJ cm$^{-2}$, above which a significantly narrowed (FWHM=1.7 nm) lasing emission was observed. Liang et al. showed multimode microcavity lasing at 78 K from mechanically exfoliated quasi-2D $(BA)_2(MA)_{n-1}Pb_nI_{3n+1}$ perovskites with a low lasing threshold of 2.6 µJ cm$^{-2}$ (Figure 14 f).[246] The lasing color was tuned by modulating the thickness of the quasi-2D halide perovskite, as shown in Figure 14 g. Recently, Qin et al. demonstrated continuous wave (CW) pumped lasing at room temperature from quasi-2D halide perovskites with a threshold of 4.7 µJ cm$^{-2}$.[256] CW lasing in layered 2D halide perovskites is difficult to achieve at room temperature, which is due to singlet-triplet exciton annihilation. However, they employed 1-naphthylmethylamine as a triplet quencher to suppress the exciton annihilation and to achieve low-threshold CW lasing from a quasi-2D halide perovskite distributed feedback cavity. In addition to the cavity lasing, random lasing from layered 2D halide perovskites have also been observed.[271] Nevertheless, defects and Auger processes that affected the efficiency of LEDs can impede the performance of quasi-2D halide perovskite lasers as well. Therefore, strategies for defect passivation and suppression of Auger recombination should be employed for utilizing layered 2D halide perovskites as efficient gain media in lasers.[253]

*6.3 Photodetectors*

Anisotropic charge carrier properties along the stacking and crystal growth directions make layered 2D halide perovskites an interesting class of photodetectors.[272–277] Anisotropy in the charge carrier properties of layered 2D halide perovskites is caused by the natural multi-quantum well structure. Therefore, the exciton diffusion occurs more efficiently towards the edges from the bulk than towards the stacking plane due to the organic barrier. In layered 2D halide perovskites, the edges of the inorganic layer contribute lower energy states.[126,278] Excitons get trapped by these states and dissociated into slowly recombining free carriers. Therefore, a large photocurrent from the edge-dissociated excitons can be detected under photoirradiation. The room-temperature carrier mobility of 5.4−7.0 cm$^2$ V$^{−1}$ s$^{−1}$ is reported at the edges of $(BA)_2(FA)_{n-1}Pb_nI_{3n+1}$ perovskites, showing that the edges are conducting.[127] However, the exact origin of such edge states in the 2D perovskite layers is not clear. While Shi et al. discussed the moisture-induced atomic rearrangement and formation of the emissive edge states in layered 2D halide perovskites,[127] Zho et al. showed that such low energy edge states are formed due to the loss of organic cation spacers at the crystal edge.[279] Although the dissociation of excitons at the edges of 2D halide perovskites were employed initially to improve the PCE solar cells,[126] it was later used to enhance the performance of photodetectors.[273,274]

The performance of a 2D halide perovskite photodetector is evaluated using various parameters such as responsivity, detectivity, and response speed.[280] Responsivity is defined as the ratio of photocurrent density to the incident light intensity. Detectivity measures the sensitivity of a photodetector above the noise/dark current. Similarly, response speed indicates how fast a photodetector reaches 90% of the maximum photocurrent after the light exposure ($\tau_{rise}$) and how fast the photocurrent decays to 10% of its maximum in dark ($\tau_{fall}$). The performance of a perovskite photodetector is affected by the bargain between the high photocurrent which requires high-quality crystals without defects and grain boundaries and a very low dark current which requires a barricade for the thermally activated charge carriers. Therefore, layered 2D halide perovskites are good candidates for photodetection due to their anisotropic charge carrier properties. Tan et al. reported $(BA)_2PbBr_4$ perovskite single crystal planar photodetector in 2016, which showed an encouraging responsivity of 2100 A W$^{-1}$ with a low dark current of $10^{-10}$ A.[281] Following this, the photodetectors based on layered 2D halide perovskites and 2D perovskite heterostructures with different compositions are known.[282–286] In later years, the performance of these photodetectors fabricated on single crystals and films was considerably improved through the modulation of quality, thickness, and orientation of the 2D crystal, interface engineering, and understanding the anisotropic charge carrier dynamics.[272–277,285,287–293] Feng et al. reported a high-performance photodetector based on nanowire-like arrays of quasi-2D $(BA)_2(MA)_{n-1}Pb_nI_{3n+1}$ [n=4] perovskites with average responsivities >1.5 × 10$^4$ A W$^{-1}$ and detectivities >7 × 10$^{15}$ Jones.[274] The organic barriers along the length of the nanowire-like array extremely suppressed the dark current to the order of ~10$^{-15}$ A, whereas the excitons diffused to the exposed edges and dissociated into the long-lived free carriers, amplifying the photocurrent. The anisotropic charge carrier properties and efficient exciton dissociation at the edges resulted in a fast response time of <30 µs, which was in the millisecond time-scale for those which did not involve the edge dissociation of excitons. Similarly, Liu et al. demonstrated a responsivity >139 A W$^{-1}$, high detectivity in the order of ~10$^{15}$ Jones, and a fast response time of <40 µs from a photodetector constructed on the (001) plane of a millimeter-sized $(PEA)_2PbI_4$ perovskite single crystal.[273] The photodetector performance based on the (001) plane was superior to (010) plane due to the high electronic conductivity along the (001) plane and the dissociation of excitons at the exposed edges which is hindered along the (010) plane due to the organic barrier. Huan et al. fabricated a photodetector on the highly oriented layered 2D halide perovskite film which showed a record fast response time of ~2.54 ns, a low dark current of $10^{-11}$ A, and very high detectivity in the order of $10^{14}$ Jones.[277] In their device, the high conductivity of excitons along the in-plane direction resulted in ultrafast response speed. Fu et al. fabricated high-performance photodetectors from the nanosheets of $C_6H_5(CH_2)_3NH_3)_3Pb_2I_7$ perovskites which show significantly low dark current (1.5 pA), fast response speed ($\tau_{rise}$=850 μs and $\tau_{fall}$=780 μs), and high detectivity (~$1.2\times10^{10}$ Jones).[294] They attributed the low dark current, fast response speed, and high detectivity to the high crystallinity and low defect density of the perovskite nanosheets. Nevertheless, the performance of photodetectors based on the layered 2D halide perovskite film is affected by the traps at the grain boundaries, which can be mitigated by preparing the highly crystalline films with large grains.[289] Also, in the case of tin-based layered 2D halide perovskite devices, self-doping due to the oxidation of $Sn^{2+}$ to $Sn^{4+}$ can increase the overall carrier conductivity and the dark current.[286] Treatment of tin-based 2D halide perovskite crystals with $SnF_2$ can improve the performance of photodetector by minimizing the self-doping through $Sn^{2+}$ vacancy passivation.[283]

*6.4 Photocatalysis*

Recently, the photocatalytic activities of 3D $ABX_3$ perovskites in water splitting and hydrogen production, $CO_2$ reduction, organic reactions, and dye degradation are explored.[11,12,295–298] Large absorption coefficient for visible light, photogeneration of free charge carriers, selective redox reaction, low cost, and facile synthesis make halide perovskites promising photocatalysts. Halide perovskite photocatalysis is demonstrated with either the pure 3D $ABX_3$ as catalyst[296] or its composite with other materials such as Pt,[297] $TiO_2$,[299] tungsten oxide,[300] reduced graphene oxide,[301] $Bi_2WO_6$,[302] and graphitic carbon nitride.[298] These composite structures help to accelerate charge separation and transport for efficient photocatalysis. Despite fulfilling most of the criteria to become an efficient photocatalyst, 3D ABX$_3$ perovskites do not survive for a long time under the conditions required for photocatalytic reactions, such as highly polar/aqueous and oxygen-rich environments and prolonged irradiation.[12] On the other hand, layered 2D halide perovskites can be used as photocatalysts with better durability.

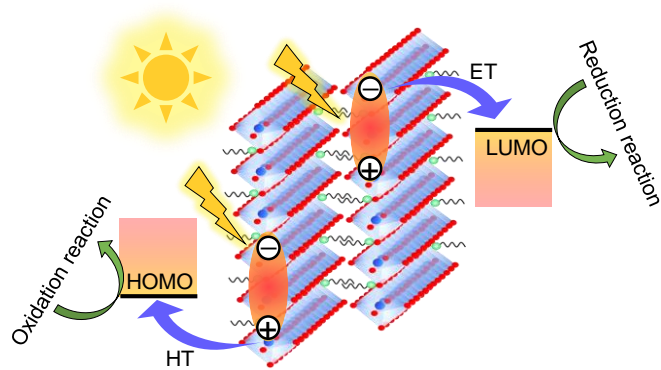

**Figure 15.** Scheme showing charge separation and photocatalytic activity of the layered 2D halide perovskites. HOMO and LUMO refer to the highest occupied molecular orbital of a hole acceptor and the lowest unoccupied molecular orbital of an electron acceptor, respectively. HT and ET are defined as the hole transfer and electron transfer processes, respectively.

The high-energy excitons generated at the wide bandgap 2D perovskite layer can easily overcome the potential energy for driving the photocatalytic reactions. Hong et al. used (HDA)$_2$BI$_4$ [HDA=hexadecylammonium, B=Pb or Sn] perovskite to catalyze a photo-redox organic synthesis involving decarboxylation and dehydrogenation reactions.[303] It involved the conversion of an indoline-2-carboxylic acid to indoline (decarboxylation) and indole (decarboxylation and dehydrogenation) in the presence of layered 2D halide perovskite as catalyst under white light illumination. They obtained a maximum yield of 98% for indoline and 84% for indole. Interestingly, there was no product formation in the absence of 2D halide perovskite photocatalyst; and even if other catalysts were used, the yield was very low, which confirmed the superior photocatalytic activity of the (HDA)$_2$BI$_4$ perovskites. Recently, Wang et al. demonstrated photocatalytic splitting of HI to generate $H_2$ by using layered 2D halide perovskites as photocatalysts.[304] They showed the dependency of $H_2$ evolution on the organic spacer cations, whereby the shortest chain length cation PMA exhibited the highest photocatalytic activity in the presence of cocatalyst Pt. This was attributed to the efficient charge transfer among the stacked 2D perovskite layers and at the perovskite-cocatalyst interface. They also claim that the conversion efficiency of ~1.57% for this reaction is the champion value among all the halide perovskite photocatalysts for $H_2$ production reported to date. Recently, Wu et al. demonstrated the photocatalytic reduction of $CO_2$ to CO with a generation rate of 21.6 $\mu mol\ g^{-1}\ h^{-1}$ by $CsPbBr_3$ perovskite nanosheets.[305] This generation rate was 3.8 times higher than the 3D $CsPbBr_3$ perovskite nanocrystals which further improved (~43.9 $\mu mol\ g^{-1}\ h^{-1}$) and became >7 times than the 3D $ABX_3$ halide perovskites when the bromide ions in the nanosheets were replaced with the mixed Br/I through the anion exchange. The excellent photocatalytic performance of these 2D halide perovskite nanosheets was attributed to the largely exposed/less-coordinated metal atoms as active sites, short charge carrier diffusion along the thickness and easy extraction of charges to $CO_2$ and $H_2O$ reactants, and enhanced visible light absorption. Nevertheless, one can argue that the strong dimensional and dielectric confinement of charge carriers imposes poor charge separation and transport in the case of layered 2D halide perovskites, which makes them poor photocatalysts. Exciton dissociation and effective charge separation in layered 2D halide perovskites can be realized by employing efficient electron/hole acceptors, and the separated charges can be utilized for photo-redox reactions, as shown schematically in Figure 15. Li et al. showed effective charge separation in strongly quantum confined $CsPbBr_3$ perovskite nanoplatelets (exciton binding energy=260 meV) by using organic molecules benzoquinone and phenothiazine as electron and hole acceptors, respectively.[306] The charge-separated states existed for >100 ns, which makes such 2D halide perovskite-organic molecule composites good photocatalyst candidates. A long-lived charge-separated state in the strongly quantum confined halide perovskites is also reported for the case of methyl viologen as an electron acceptor.[307]

## 7. Summary and Outlook

Layered 2D halide perovskites are an interesting class of semiconductor materials in which the optoelectronic properties are tuned by the dimensional and dielectric confinements. They show better environmental stability compared to the 3D $ABX_3$ analogs, owing to their unique structure which comprises bulky hydrophobic organic ligands to become a part of the crystal structure. High PL QY in 2D or quasi-2D halide perovskites along with the processes such as efficient energy transfer between the inorganic layers with different thickness, exciton dissociation at the edges, and charge separation can be utilized for efficient solar to electrical energy conversion, solar to chemical energy conversion, light-emitting applications, and photodetectors. The inclusion of charge-transfer complexes between the stacked perovskite layers could be an efficient strategy to separate strongly bound excitons in layered 2D halide perovskites. Nevertheless, challenges remain in stabilizing the colloidal 2D halide perovskite nanoplatelets and nanosheets during the synthesis, purification, and long-term storage. Polymer or inorganic coatings can improve the moisture- and photostability of colloidal 2D perovskite nanoplatelets and nanosheets. The selective crystallization of layered 2D halide perovskites with higher n is often limited by the precipitation of monolayered structures due to their low enthalpy of formation.[51] Such large bandgap 2D halide perovskite monolayers show enhanced nonradiative recombination due to the efficient exciton-phonon coupling, Auger processes, and defects. The nonradiative recombination in these materials can be suppressed through the passivation of defects and Auger recombination centers by solvent engineering and inorganic salt treatment. Further, the anisotropic charge carrier properties require specific orientation and interfacial engineering of the 2D perovskite crystals for efficient carrier transport and extraction in different optoelectronic applications. 2D/3D perovskite heterostructures, on the other hand, can show better carrier mobility due to their suitable band alignment. Although the broadband emission observed in some layered 2D halide perovskites has shown promise towards the white-emitting LEDs, insufficient understanding of the exact origin of such phenomenon has limited further development. Therefore, full-fledged utilization of layered 2D halide perovskites for diverse optoelectronic applications requires better strategies to control the growth and stability of the crystals in colloidal and noncolloidal states and an advanced understanding of the various photophysical and chemical processes therein.

**Conflicts of Interest**

The authors declare no conflicts of interest.

**Acknowledgments**

SG acknowledges Alexander von Humboldt (AvH) Foundation for the postdoctoral research fellowship.